\documentclass[journal,onecolumn,12pt, romanappendices]{IEEEtran-v17}
\pdfoutput=1

\usepackage{amsmath}
\usepackage{amsthm}
\usepackage{amssymb}
\usepackage{footnote}
\usepackage{xcolor}
\usepackage[nosort]{cite}
\usepackage[british]{babel}
\usepackage{tikz}

\usepackage{ifthen}

\newboolean{conf}
\setboolean{conf}{false} 

\usepackage{bm}
\usepackage{upgreek}
\usepackage{amssymb}
\usepackage{amsfonts}
\usepackage{mathrsfs}
\usepackage{xspace}

\makeatletter
\def\amsbb{\use@mathgroup \M@U \symAMSb}
\makeatother

\newcommand{\lefto}{\mathopen{}\left}
\newcommand{\safemath}[2]{\newcommand{#1}{\ensuremath{#2}\xspace}}

\safemath{\opE}{\amsbb{E}}
\newcommand{\Ex}[2]{\ensuremath{\amsbb{E}_{#1}\mathopen{}\left[#2\right]}} 	
\safemath{\prob}{\amsbb{P}}
\safemath{\pp}{\mathrm{Pr}}

\safemath{\define}{\triangleq}			
\safemath{\gradient}{\nabla}
\safemath{\bigO}{\mathcal{O}}
\safemath{\littleo}{\mathit{o}}

\safemath{\Real}{\mathrm{Re}} 

\safemath{\diag}{\mathrm{diag}}

\safemath{\jpg}{\mathcal{CN}}			
\safemath{\complexset}{\amsbb{C}}
\safemath{\realset}{\amsbb{R}}
\safemath{\extendreal}{\overline{\realset}}
\safemath{\natunum}{\mathbb{N}}
\safemath{\posrealset}{\realset_{+}}
\safemath{\integerset}{\amsbb{N}}
\newcommand{\given}{\vert}				

\safemath{\spanm}{\mathrm{span}}

\newtheorem{thm}{Theorem}
\newtheorem{lemma}[thm]{Lemma}
\newtheorem{dfn}{Definition}
\newtheorem{rem}{Remark}

\newtheorem{cor}[thm]{Corollary}

\newcommand{\indfun}[1]{\mathbf{1}\mathopen{}\left\{#1\right\}}

\safemath{\matA}{\mathsf{A}}
\safemath{\matB}{\mathsf{B}}
\safemath{\matC}{\mathsf{C}}
\safemath{\matD}{\mathsf{D}}
\safemath{\matE}{\mathsf{E}}
\safemath{\matF}{\mathsf{F}}
\safemath{\matG}{\mathsf{G}}
\safemath{\matH}{\mathsf{H}}
\safemath{\matI}{\mathsf{I}}
\safemath{\matJ}{\mathsf{J}}
\safemath{\matK}{\mathsf{K}}
\safemath{\matL}{\mathsf{L}}
\safemath{\matM}{\mathsf{M}}
\safemath{\matN}{\mathsf{N}}
\safemath{\matO}{\mathsf{O}}
\safemath{\matP}{\mathsf{P}}
\safemath{\matQ}{\mathsf{Q}}
\safemath{\matR}{\mathsf{R}}
\safemath{\matS}{\mathsf{S}}
\safemath{\matT}{\mathsf{T}}
\safemath{\matU}{\mathsf{U}}
\safemath{\matV}{\mathsf{V}}
\safemath{\matW}{\mathsf{W}}
\safemath{\matX}{\mathsf{X}}
\safemath{\matY}{\mathsf{Y}}
\safemath{\matZ}{\mathsf{Z}}
\safemath{\matSigma}{\mathsf{\Sigma}}
\safemath{\matPhi}{\mathsf{\Phi}}
\safemath{\matLambda}{\mathsf{\Lambda}}
\safemath{\matDelta}{\mathsf{\Delta}}

\safemath{\randveca}{\bm{A}}
\safemath{\randvecb}{\bm{B}}
\safemath{\randvecc}{\bm{C}}
\safemath{\randvecd}{\bm{D}}
\safemath{\randvece}{\bm{E}}
\safemath{\randvecf}{\bm{F}}
\safemath{\randvecg}{\bm{G}}
\safemath{\randvech}{\bm{H}}
\safemath{\randveci}{\bm{I}}
\safemath{\randvecj}{\bm{J}}
\safemath{\randveck}{\bm{K}}
\safemath{\randvecl}{\bm{L}}
\safemath{\randvecm}{\bm{M}}
\safemath{\randvecn}{\bm{N}}
\safemath{\randveco}{\bm{O}}
\safemath{\randvecp}{\bm{P}}
\safemath{\randvecq}{\bm{Q}}
\safemath{\randvecr}{\bm{R}}
\safemath{\randvecs}{\bm{S}}
\safemath{\randvect}{\bm{T}}
\safemath{\randvecu}{\bm{U}}
\safemath{\randvecv}{\bm{V}}
\safemath{\randvecw}{\bm{W}}
\safemath{\randvecx}{\bm{X}}
\safemath{\randvecy}{\bm{Y}}
\safemath{\randvecz}{\bm{Z}}
\safemath{\randvecLambda}{\bm{\Lambda}}

\safemath{\randmatA}{\amsbb{A}}
\safemath{\randmatB}{\amsbb{B}}
\safemath{\randmatC}{\amsbb{C}}
\safemath{\randmatD}{\amsbb{D}}
\safemath{\randmatE}{\amsbb{E}}
\safemath{\randmatF}{\amsbb{F}}
\safemath{\randmatG}{\amsbb{G}}
\safemath{\randmatH}{\amsbb{H}}
\safemath{\randmatI}{\amsbb{I}}
\safemath{\randmatJ}{\amsbb{J}}
\safemath{\randmatK}{\amsbb{K}}
\safemath{\randmatL}{\amsbb{L}}
\safemath{\randmatM}{\amsbb{M}}
\safemath{\randmatN}{\amsbb{N}}
\safemath{\randmatO}{\amsbb{O}}
\safemath{\randmatP}{\amsbb{P}}
\safemath{\randmatQ}{\amsbb{Q}}
\safemath{\randmatR}{\amsbb{R}}
\safemath{\randmatS}{\amsbb{S}}
\safemath{\randmatT}{\amsbb{T}}
\safemath{\randmatU}{\amsbb{U}}
\safemath{\randmatV}{\amsbb{V}}
\safemath{\randmatW}{\amsbb{W}}
\safemath{\randmatX}{\amsbb{X}}
\safemath{\randmatY}{\amsbb{Y}}
\safemath{\randmatZ}{\amsbb{Z}}
\safemath{\randmatSigma}{\mathbb{\Sigma}}
\safemath{\randmatPhi}{\mathbb{\Phi}}

\safemath{\veca}{\bm{a}}
\safemath{\vecb}{\bm{b}}
\safemath{\vecc}{\bm{c}}
\safemath{\vecd}{\bm{d}}
\safemath{\vece}{\bm{e}}
\safemath{\vecf}{\bm{f}}
\safemath{\vecg}{\bm{g}}
\safemath{\vech}{\bm{h}}
\safemath{\veci}{\bm{i}}
\safemath{\vecj}{\bm{j}}
\safemath{\veck}{\bm{k}}
\safemath{\vecl}{\bm{l}}
\safemath{\vecm}{\bm{m}}
\safemath{\vecn}{\bm{n}}
\safemath{\veco}{\bm{o}}
\safemath{\vecp}{\bm{p}}
\safemath{\vecq}{\bm{q}}
\safemath{\vecr}{\bm{r}}
\safemath{\vecs}{\bm{s}}
\safemath{\vect}{\bm{t}}
\safemath{\vecu}{\bm{u}}
\safemath{\vecv}{\bm{v}}
\safemath{\vecw}{\bm{w}}
\safemath{\vecx}{\bm{x}}
\safemath{\vecy}{\bm{y}}
\safemath{\vecz}{\bm{z}}
\safemath{\veclambda}{\bm{\lambda}}
\safemath{\vecpi}{\bm{\pi}}
\safemath{\vecsigma}{\bm\sigma}              			
\safemath{\veczero}{\mathbf{0}} 

\safemath{\setA}{\mathcal{A}}
\safemath{\setB}{\mathcal{B}}
\safemath{\setC}{\mathcal{C}}
\safemath{\setD}{\mathcal{D}}
\safemath{\setE}{\mathcal{E}}
\safemath{\setF}{\mathcal{F}}
\safemath{\setG}{\mathcal{G}}
\safemath{\setH}{\mathcal{H}}
\safemath{\setI}{\mathcal{I}}
\safemath{\setJ}{\mathcal{J}}
\safemath{\setK}{\mathcal{K}}
\safemath{\setL}{\mathcal{L}}
\safemath{\setM}{\mathcal{M}}
\safemath{\setN}{\mathcal{N}}
\safemath{\setO}{\mathcal{O}}
\safemath{\setP}{\mathcal{P}}
\safemath{\setQ}{\mathcal{Q}}
\safemath{\setR}{\mathcal{R}}
\safemath{\setS}{\mathcal{S}}
\safemath{\setT}{\mathcal{T}}
\safemath{\setU}{\mathcal{U}}
\safemath{\setV}{\mathcal{V}}
\safemath{\setW}{\mathcal{W}}
\safemath{\setX}{\mathcal{X}}
\safemath{\setY}{\mathcal{Y}}
\safemath{\setZ}{\mathcal{Z}}
\safemath{\emptySet}{\varnothing}

\newcommand{\error}{\epsilon}
\newcommand{\secrecy}{\delta}
\newcommand{\tvar}{d}

\newcommand{\msg}{W}
\newcommand{\CS}{C_{\mathrm{S}}}
\newcommand{\uniformX}{P_{X}^{\mathrm{unif}}}

\begin{document}

\IEEEoverridecommandlockouts
 
\title{Finite-Blocklength Bounds for Wiretap Channels}

\author{\IEEEauthorblockN{Wei Yang$^1$, Rafael F. Schaefer$^{2}$, and H. Vincent Poor$^1$}
\thanks{The work of H. V. Poor and W. Yang was supported in part by the US  National Science Foundation under Grants  CCF-1420575 and ECCS-1343210.
 The work of R. F. Schaefer was supported by the German Research Foundation (DFG) under Grant WY 151/2-1.}\\
\IEEEauthorblockA{
$^1$Princeton University, Princeton, NJ, 08544 USA\\
$^2$Technische Universit\"{a}t Berlin, 10587 Berlin, Germany}
}
\maketitle

\begin{abstract}
This paper investigates the maximal  secrecy rate over a wiretap channel subject to reliability and secrecy constraints at a given blocklength. 
New achievability and converse bounds are derived, which are shown to be tighter than existing bounds. The bounds also lead to the tightest second-order coding rate for discrete memoryless and Gaussian wiretap channels. 
\end{abstract}

\section{Introduction}

We consider the problem of secure communication over a wiretap channel  $(\setX,  P_{YZ\given X}, \setY\times\setZ)$ with a transmitter $X$ (Alice), a legitimate receiver $Y$ (Bob), and an eavesdropper $Z$ (Eve).
The transmitter aims to communicate  a message $W$ to the receiver while keeping it secret from the eavesdropper. 
We are interested in finding upper and lower bounds on the maximal secret communication rate $R^*(n,\error,\secrecy)$ for a given blocklength $n$, error probability $\error$, and information leakage~$\delta$. Throughout the paper, the information leakage to the eavesdropper is measured by the total variation distance between $P_{WZ}$ and $P_{W}P_Z$.

The wiretap channel model was originally introduced by Wyner~\cite{wyner1975-10a} (using a different metric for the information leakage). Wyner established that, in the asymptotic limit of $\delta \to 0$, $\error\to 0$, and $n\to\infty$, the maximal secrecy rate for a degraded discrete memoryless wiretap channel (DM-WTC) converges to the secrecy capacity 
\begin{IEEEeqnarray}{rCl}
\CS  = \max\limits_{P_X} \mathopen{}\Big( I(X;Y) - I(X;Z)\Big).
\label{eq:secrecy-capacity-intro}
\end{IEEEeqnarray}
This result was later generalized by Csisz\'{a}r and K\"{o}rner to general discrete memoryless channels~\cite{csiszar1978-05a} and by Leung-Yan-Cheong and Hellman to Gaussian wiretap channels~\cite{leung-yan-cheong1978-07a}.

Nonasymptotically, bounds and approximations for $R^*(n,\error,\delta)$ have been developed.  Hayashi~\cite{hayashi2006-04a} established general achievability bounds using channel resolvability~\cite{han93-03} and studied the secrecy exponent (i.e., the exponential decreasing rate of the information leakage) for a fixed communication rate.  Later, he improved the secrecy exponent by leveraging the \emph{privacy amplification} technique~\cite{hayashi13-11a}. For the setting of fixed $\error$ and $\delta$ and $n\to\infty$, Yassaee et al.~\cite{yassaee2013-07a} derived an achievability bound  on the second-order coding rate~\cite{hayashi2009-11a} (also known as dispersion~\cite{polyanskiy10-05}) by using the output statistics of random binning, which improves an earlier result by Tan~\cite{tan2012-11a}.
\ifthenelse{\boolean{conf}}{}{
 Achievability bounds for wiretap channels with other information leakage metrics can be found in~\cite{hayashi2011-06a,parizi2015-10a, yassaee2015-06a}.}
On the converse side,  Tyagi and Watanabe proposed a one-shot converse bound for the problem of secret key agreement~\cite{tyagi2015-05a}, which exploits  hypothesis testing. Building upon the technique in~\cite{tyagi2015-05a}, Hayashi et al.~\cite{hayashi2014-09a} established a converse bound for wiretap channels, which leads to the strong converse for the degraded case (see also~\cite{tan2015-09a}). 

\subsubsection*{Contributions}
In this paper, we propose new achievability and converse bounds on $R^*(n,\error,\delta)$ for general wiretap channels. Our achievability bound is based on a  new privacy amplification lemma, which refines the results in~\cite{hayashi13-11a,watanabe2013-07a}. Our converse bound  is motivated by~\cite{tyagi2015-05a,hayashi2014-09a}, and relates secrecy communication to  binary hypothesis testing. 
The bounds are computed for a Gaussian wiretap channel. In this case, both our converse and achievability bounds are uniformly tighter than the best existing bounds (to the best of our knowledge).  

By analyzing the behavior of our bounds in the regime of fixed~$\error$ and $\delta$ and $n\to\infty$, we obtain upper and lower bounds on the second-order coding rate of DM-WTCs and Gaussian wiretap channels. Our achievable second-order coding rate is tighter than the ones in~\cite{yassaee2013-07a,tan2012-11a}, hence showing the advantage of privacy amplification in the finite-blocklength regime, among all constructions that are secrecy-capacity achieving.  

We also derive two converse bounds for a specific class of codes, called \emph{partition codes}, which are obtained by  partitioning ordinary (nonsecret) codes for the  channel from the transmitter to the legitimate receiver.  Our bounds reveal an interesting connection between secure communication and list decoding at the eavesdropper.

\subsubsection*{Notation}
The cardinality of a set $\setA$ is denoted by $|\setA|$.
For an input distribution $P_{X}$ and a random transformation $P_{Y|X}$, we let $P_{Y|X}\circ P_X$ denote the marginal distribution of $P_{X}P_{Y|X}$ on $Y$.
We shall consider the following metrics between two probability distributions $P$ and $Q$ a common space~$\setA$:
\begin{itemize}
\item  $\ell_1$ distance
\begin{equation}
\|P-Q\|_1 \define \sum\limits_{x\in\setA} |P(x)-Q(x)|
\end{equation}
\item  total variation distance
\begin{equation} 
\tvar(P,Q) \define \sup_{\setE\subset \setA} |P(\setE) - Q(\setE)| = \frac{1}{2}\|P-Q\|_1
\label{eq:def-total-var}
\end{equation}
\item $E_{\gamma}$ metric~\cite{liu2015-11a}
\begin{IEEEeqnarray}{rCl}
E_{\gamma}(P,Q) &\define& P\lefto[\frac{dP}{dQ} \geq \gamma\right] - \gamma Q\lefto[\frac{dP}{dQ} \geq \gamma\right] \label{eq:def-e-gamma-1}\\
&=& \sup_{\setE \subset \setA} P[\setE] -\gamma Q[\setE]. \label{eq:def-e-gamma-2}
\end{IEEEeqnarray}
\end{itemize}
Furthermore, we define a randomized test between $P$ and $Q$ as a random transformation $P_{T\given X}: \setA\mapsto\{0,1\}$ where $0$ indicates that the test chooses $Q$.
We shall need the following performance metric for the test between~$P$ and~$Q$:
\begin{IEEEeqnarray}{rCl}
\label{eq:def-beta}
\beta_\alpha(P,Q) \define \min\int P_{T\given X}(1\given x)  Q(d x)
\end{IEEEeqnarray}
where the minimum is over all  tests $P_{T\given X}$
satisfying
\begin{IEEEeqnarray}{rCl}
 \int P_{T\given X} (1\given x) P(dx)\geq \alpha.
\end{IEEEeqnarray}

\section{Channel Model and Secrecy Codes}
We consider the wiretap channel model introduced by Wyner~\cite{wyner1975-10a}, which is denoted by the tuple $(\setX, P_{YZ \given X}, \setY\times\setZ)$. 
A secrecy code for the wiretap channel  is defined as follows.

\begin{dfn} An $(M,\error,\delta)$ secrecy code for the wiretap channel $(\setX, P_{YZ \given X}, \setY\times\setZ)$ consists of 
\begin{itemize}
\item a message $\msg$ which is equiprobable on the set  $\setM \define \{1,\ldots,M\}$,
\item a \emph{randomized} encoder that generates a codeword $X(m)$, $m\in\setM$ according to a conditional probability distribution $P_{X \given W =m}$, and
\item a decoder $g: \setY \to \setM $ that assigns an estimate $\hat{W}$ to each received signal $Y \in\setY$.
 \end{itemize}
 Furthermore, the encoder and decoder satisfy the average error probability constraint 
\begin{equation}
\prob[g(Y) \neq W ] \leq \error
\label{eq:error-constraint-wtc}
\end{equation}
where $Y\sim P_{Y|W} \define P_{Y|X}\circ P_{X|W}$, and the secrecy constraint 
\begin{equation}
\tvar(P_{WZ}, P_{W }P_Z) \leq \secrecy.
\label{eq:def-secrecy-constraint-def1}
\end{equation}
 \end{dfn}
 %

 An $(M,\error,\delta)$ secrecy code for the channel $(\setX^n,  P_{Y^nZ^n \given X^n} ,     \setY^n\times\setZ^n)$ will be called an $(n,M,\error,\secrecy)$ secrecy code.  Furthermore, the maximal secrecy rate is defined as
\begin{IEEEeqnarray}{rCl}
R^*(n,\error,\secrecy) \define \max\lefto\{\frac{\log M}{n}\!:\, \exists (n, M,\error,\secrecy)\,\text{secrecy code} \right\}. \IEEEeqnarraynumspace
\end{IEEEeqnarray}

We shall also consider a class of codes which are constructed by partitioning good channel codes for the legitimate channel $P_{Y|X}$. 
\begin{dfn}[Partition codes] 
\label{dfn:partition-codes}

An $(M,\error,\delta)$ partition code for the wiretap channel $(\setX, P_{YZ \given X}, \setY\times\setZ)$ is a tuple $(\setC,\pi, P_{X|W})$, where
\begin{itemize}
\item $\setC=\{x_1,\ldots,x_N\}$ with decoder $g_{\mathrm{le}}: \setY \to \setC$ is a channel code for the legitimate channel $P_{Y|X}$;
\item  $\pi : \setC \mapsto \setM$ is a function that partitions $\setC$ into $M\leq N$ disjoint subsets $\setC = \cup_{m=1}^{M} \pi^{-1}(m)$;
\item and $P_{X|W}$ is a stochastic encoder with $P_{X|W=m}$, $m\in\setM$, supported on $\pi^{-1}(m)$. 
\end{itemize}
Furthermore, the encoder and decoder satisfy  the secrecy constraint~\eqref{eq:def-secrecy-constraint-def1} and the average  error probability constraint
\begin{IEEEeqnarray}{rCl}
\prob[ g_{\mathrm{le}}(Y) \neq X ] \leq \error 
\end{IEEEeqnarray}
where $X \sim P_{X |W} \circ P_{W}$ with $W$ equiprobable over $\setM$, and $Y\sim P_{Y|X}$.
\end{dfn}

Note that the class of partition codes includes most of the existing coding  schemes   for wiretap channels (e.g.,~\cite{thangaraj2007-08a,mahdavifar2011-10a,harrison2013-09a}).

\begin{dfn}[Uniform-partition codes] 
\label{dfn:unif-partition-codes}
 An $(M,\error,\delta)$ partition code $(\setC, \pi, P_{X|W})$ is called a uniform-partition code if $|\pi^{-1}(m)|$ does not depend on $m$ and if $P_{X|W=m}$ is the uniform distribution on $\pi^{-1}(m)$ for every $m\in\setM$.
\end{dfn}

The partition codes and uniform-partition codes introduced above are practically appealing, because  they can reuse the decoder of the original  channel code. Furthermore, the encoder of a uniform-partition code can be obtained by concatenating the encoder of the original channel code with a uniform random number generator.

\section{Main Results}

\subsection{Achievability Bound}
The following lemma builds upon the leftover hash lemma  (see, e.g.,~\cite[Lemma~5.4.3]{renner2005-09a}) and refines the result in~\cite[Cor.~2]{watanabe2013-07a}.
\ifthenelse{\boolean{conf}}{
 Due to space constraints, we have omitted the proofs of some results. They can be found in~\cite{yang16-wtp}.}
{
}

\begin{lemma}
\label{lemma:privacy-amplification}
Let $\setC=\{x_1,\ldots,x_{KM}\}$ be an arbitrary codebook of cardinality $KM$ with $K,M \in \natunum$. Let $P_X$ be the uniform distribution over $\setC$, and let $P_{Z} \define P_{Z|X} \circ P_{X}$.
There exists a function $\pi: \setC \to \setM$ such that $\pi(X)$ is equiprobable over $\setM$, and that for every $\gamma>0$ and every $Q_Z$
\begin{IEEEeqnarray}{rCl}
&& d(P_{\pi(X)Z} , P_{\pi(X)}P_Z ) \leq 
E_{\gamma}(P_{XZ},P_XQ_Z)  \notag\\
&&\qquad\qquad\qquad + \, \frac{1}{2}\sqrt{\frac{\gamma}{K} \Ex{}{\exp\lefto( - \big|\imath(X;Z)  - \log\gamma \big| \right)} } \IEEEeqnarraynumspace
\label{eq:privacy-amplification-lemma}
\end{IEEEeqnarray}
where 
\begin{IEEEeqnarray}{rCl}
\imath(x;z) \define \log \frac{dP_{Z|X=x}}{dQ_Z}(z)
\label{eq:info-den-qz}
\end{IEEEeqnarray}
and the expectation is taken with respect to $(X,Z)\sim P_{XZ}$. 
\end{lemma}
\ifthenelse{\boolean{conf}}{
}{
\begin{IEEEproof}
See Appendix~\ref{app:proof-of-lemma-privacy-amp}.
\end{IEEEproof}}
\begin{rem}
The bound~\cite[Cor.~2]{watanabe2013-07a} can be obtained from~\eqref{eq:privacy-amplification-lemma} by upper-bounding the expectation term on the right-hand side (RHS) of~\eqref{eq:privacy-amplification-lemma} by $1$ and by using~\cite[Lem.~19]{hayashi2014-11a}. This implies that~\eqref{eq:privacy-amplification-lemma} is stronger than~\cite[Cor.~2]{watanabe2013-07a}.
\end{rem}

Lemma~\ref{lemma:privacy-amplification} implies that we can convert an arbitrary (nonsecret) channel code for the legitimate channel $P_{Y|X}$ into a secrecy code with bounded information leakage. 
This step is commonly referred to as privacy amplification (see, e.g.,~\cite[p.~413]{Csiszar11}). 
By combining the channel coding  achievability bounds in~\cite{polyanskiy10-05} with Lemma~\ref{lemma:privacy-amplification}, we obtain the following achievability bound for a wiretap channel. 

\begin{thm}
\label{thm:ach-wiretap}
Let $P_X$ be a probability distribution supported on $\setA \subset \setX$. For every $K\in \natunum$,  every $\gamma>0$,  and every $Q_Z$,  there exists an $(M,\error,\secrecy)$ uniform-partition code for the wiretap channel $(\setX, P_{YZ \given X},\setY\times\setZ)$ that satisfies
\begin{IEEEeqnarray}{rCl}
\delta
  &\leq&  
  \sup_{x\in \setA}  E_{\gamma}(P_{Z|X=x},Q_Z) \notag\\
  && +\, \sup_{x\in \setA} \frac{1}{2}\sqrt{\frac{\gamma}{K} \Ex{P_{Z|X=x}}{ \exp\lefto( - \big|\imath(x;Z)  - \log\gamma \big| \right)}  } 
 \IEEEeqnarraynumspace
\label{eq:thm-secrecy-bound}
\end{IEEEeqnarray}
and that
\begin{IEEEeqnarray}{rCl}
\error &\leq&   \min\mathopen{}\Big\{ \error_{\mathrm{RCU}}(MK) , \error_{\mathrm{DT}}\mathopen{}\Big(\frac{MK-1}{2}\Big)\Big\} \label{eq:min-rcu-beta-bound}.
\end{IEEEeqnarray}
Here, $\imath(x;z)$ is defined in~\eqref{eq:info-den-qz}, 
\begin{equation} 
\error_{\mathrm{RCU}}(a)  \define \Ex{}{ \min\{1, (a-1)\prob[  i(\bar{X};Y) \geq i(X;Y) \given X, Y]\}} 
\label{eq:def-RCU} 
\end{equation}
where $i(x;y)\define \log \frac{dP_{Y|X}(y|x)}{dP_Y(y)}$,  $(X,Y,\bar{X}) \sim P_{XY}P_{X}$, and 
\begin{IEEEeqnarray}{rCl}
\epsilon_{\mathrm{DT}} (a) \define 1- E_{a}(P_{XY},P_XP_Y).
\end{IEEEeqnarray}

\end{thm}
\ifthenelse{\boolean{conf}}{}{
\begin{IEEEproof}
See Appendix~\ref{app:proof-ach-thm}.
 \end{IEEEproof}}

\subsection{Converse Bounds for General Secrecy Codes}

We first present a general converse bound.
\begin{thm}
Every $(M,\error,\secrecy)$ secrecy code satisfies 
\begin{equation}
M \leq \sup\limits_{P_{X|W}} 
\inf_{0<\tau<1-\secrecy} \inf_{Q_Y}\frac{\beta_{\secrecy+\tau} (P_{WZ},P_WP_Z )}{\tau \beta_{1-\error}(P_{WY} , P_WQ_Y)} \IEEEeqnarraynumspace
\label{eq:thm-general-converse}
\end{equation}
where $W$ is equiprobable over $\setM$, and $P_Y,P_Z$ are the marginal distributions of the Markov chains $W\to X \to Y$ and $W \to X\to Z$, respectively.
\end{thm}
\begin{IEEEproof}
By the meta-converse bound~\cite{polyanskiy10-05} for channel coding,   every $(M,\error,\secrecy)$ secrecy code satisfies 
\begin{equation}
M \leq  \inf\limits_{Q_{Y}} \frac{1}{\beta_{1-\error}(P_{\msg Y} , P_{\msg} Q_Y) }.
\label{eq:meta-converse-bound}
\end{equation}
Furthermore, by the secrecy constraint,  
\begin{IEEEeqnarray}{rCl}
\secrecy \geq d(P_{\msg Z}, P_{\msg} P_Z) \geq \secrecy +  \tau - \beta_{\secrecy +\tau} (P_{\msg Z} , P_{\msg} P_Z ) \IEEEeqnarraynumspace
\label{eq: secrecy-constraint-alpha-beta}
\end{IEEEeqnarray}
where the last step follows from~\eqref{eq:def-total-var}. 
Rearranging the terms in~\eqref{eq: secrecy-constraint-alpha-beta}, we conclude that 
\begin{equation}
\beta_{\secrecy +\tau } (P_{\msg Z}, P_{\msg} P_Z ) \geq \tau.
\label{eq:beta-tau-geq-tau}
\end{equation}
Combining~\eqref{eq:beta-tau-geq-tau}  with~\eqref{eq:meta-converse-bound}, and optimizing the resulting bound over all stochastic encoders $P_{X|W}$,  we obtain~\eqref{eq:thm-general-converse}.
\end{IEEEproof}

The bound~\eqref{eq:thm-general-converse} is in general difficult to compute or analyze. 
Next, we prove a converse bound, which is motivated by the converse  in~\cite{tyagi2015-05a} for secret key generation. This bound is both numerically and analytically  tractable. 
\begin{thm}
\label{thm:converse-wei}
Let $Q_{Y|Z}: \setY\to\setZ$ be an arbitrary random transformation. Then, every $(M,\error,\secrecy)$ secrecy code for the wiretap channel  $(\setX, P_{YZ|X}, \setY\times\setZ)$ satisfies 
\begin{IEEEeqnarray}{rCl}
M\leq \inf_{\tau \in(0, 1-\error -\secrecy) } \frac{\tau +\secrecy}{\tau \beta_{1-\error-\secrecy -\tau} (P_{XYZ}, P_{XZ}Q_{Y|Z}) } \IEEEeqnarraynumspace
\label{eq:converse-bound-general}
\end{IEEEeqnarray} 
where $P_{XYZ}$ denotes the distribution induced by the code. 
\end{thm}
\ifthenelse{
\boolean{conf}}{}{
\begin{IEEEproof}
See Appendix~\ref{app:proof-thm-general-converse}.
\end{IEEEproof}
}
\begin{rem}
Using the result in~\cite{tyagi2015-05a}, Hayashi, Tyagi, and Watanabe recently derived the following converse bound~\cite{hayashi2014-09a} 
\begin{equation}
M \leq \inf_{\tau \in(0, 1-\error-\secrecy)} \frac{1}{\tau^2  \beta_{1-\error-\secrecy-\tau} (P_{XYZ}, P_{XZ}Q_{Y|Z}) } .
\label{eq:hayashi-bound-wtc}
\end{equation}
Our bound is stronger than~\eqref{eq:hayashi-bound-wtc} since $(\tau+\secrecy)/\tau < 1/\tau^2$.

\end{rem}

\subsection{Converse Bounds for Partition Codes}

In this section, we develop two converse bounds for partition codes.  The first bound is based on the following converse result for privacy amplification.

\begin{lemma}
\label{thm:converse-partition-egamma}
Consider an $(M,\error,\delta)$ partition code  $(\setC, \pi, P_{X|W} )$. 
Let $P_{WXZ} $ be the distribution defined by the Markov chain $W \to X\to Z$ where $W$ is equiprobable over $\setM$. Then, for every~$Q_{Z} $  we have
\begin{IEEEeqnarray}{rCl}
d(P_{WZ}, P_WQ_Z) &\geq&  E_{N/M}( P_{XZ} , \uniformX Q_Z) 
\label{eq:thm-converse-pa}
\end{IEEEeqnarray}
where $N\define |\setC|$ and $\uniformX$ is the uniform distribution over~$\setC$.
\end{lemma}
\ifthenelse{\boolean{conf}}{}{
\begin{IEEEproof}
See Appendix~\ref{app:proof-pa-converse-egamma}.
\end{IEEEproof}}

As a corollary of Lemma~\ref{thm:converse-partition-egamma}, we obtain the following  converse bound for channel resolvability.

\begin{cor}
For every $\setC=\{x_1,...,x_N\}$, every $P_{Y|X}$, and every $Q_{Y}$, we have 
\begin{equation}
\label{eq:converse-resolvability}
d(P_{Y\given \setC}, Q_Y) \geq E_{N}( \uniformX P_{Y\given X}, \uniformX Q_{Y} )
\end{equation}
where $\uniformX$ denotes the uniform distribution over $\setC$, and $P_{Y\given \setC} \define P_{Y|X} \circ \uniformX$. 
\end{cor}

By combining~\eqref{eq:thm-converse-pa} with the meta-converse bound on channel coding~\cite[Th.~26]{polyanskiy10-05}, we obtain the following converse bound for partition codes.

\begin{thm}
\label{thm:partition-converse-beta-beta}
Consider an  $(M,\error,\delta)$ partition code $(\setC, \pi, P_{X|W})$. 
Let $P_{XY} $ and $P_{XZ}$ be the distributions defined by the Markov chain $W \to X\to (Y,Z)$ where $W$ is equiprobable over $\setM$. 
Then, 
\begin{IEEEeqnarray}{rCl}
M\leq \inf\limits_{Q_Y} \inf\limits_{\tau \in (0,1-\delta)} \frac{\beta_{\delta +\tau }(P_{XZ} , \uniformX P_Z)  }{\tau \beta_{1-\error} (P_{XY}, \uniformX Q_Y)} 
\label{eq:thm-conv-part-code}
\end{IEEEeqnarray}
where   $\uniformX $ denotes the uniform distribution over $\setC$. 
Furthermore, if the code is an uniform-partition code, then 
\begin{IEEEeqnarray}{rCl}
M\leq \sup\limits_{P_X} \inf\limits_{Q_Y}  \inf\limits_{\tau \in (0,1-\delta)} \frac{\beta_{\delta +\tau }(P_{XZ} , P_X P_Z)  }{\tau \beta_{1-\error} (P_{XY}, P_XQ_Y)} .
\label{eq:thm-conv-eq-part-code}
\end{IEEEeqnarray}
\end{thm}
\begin{IEEEproof}
By Lemma~\ref{thm:converse-partition-egamma}, every $(M,\error,\delta)$ partition code satisfies
\begin{IEEEeqnarray}{rCl}
\delta &\geq& E_{N/M} (P_{XZ}, \uniformX P_Z)\\
 &\geq& \delta+\tau -\frac{N}{M}\beta_{\delta+\tau}(P_{XZ}, \uniformX P_Z).
 \label{eq:conv-egamma-part-delta}
\end{IEEEeqnarray}
Since the error probability of the code $\setC$ is upper-bounded by~$\error$, by the meta-converse theorem~\cite[Th.~26]{polyanskiy10-05} and~\cite{vazquezvilar15-10}, 
\begin{equation}
\frac{1}{N} \geq \sup\limits_{Q_Y}\beta_{1-\error}(P_{XY},\uniformX Q_Y).
\label{eq:meta-conv-part}
\end{equation}
Substituting~\eqref{eq:meta-conv-part} into~\eqref{eq:conv-egamma-part-delta} and optimizing over $\tau$ we conclude~\eqref{eq:thm-conv-part-code}. The bound~\eqref{eq:thm-conv-eq-part-code} follows from~\eqref{eq:thm-conv-part-code} by observing that, for uniform-partition codes, $P_X = \uniformX$. 
\end{IEEEproof}

The next bound relates the secrecy $\delta$ of a partition code to  the error probability of the list decoding at the eavesdropper. To state our result, we first give some definitions. Consider an arbitrary partition code $(\setC, \pi, P_{X|W})$. Suppose that the eavesdropper attempts to perform list decoding for the transmitted codeword $X$. More specifically, upon reception of a signal~$Z$, the eavesdropper outputs a list $\setL(Z)\subset \setC$ of codewords. The performance  of the eavesdropper's list decoding is measured by the maximum list size 
\begin{equation}
L\define \max_{z\in\setZ} |\setL(z)|
\label{eq:list-size-eve}
\end{equation}
and the   error probability
\begin{IEEEeqnarray}{rCl}
\error_{\mathrm{ld}} \define P_{XZ}[ X\notin \setL(Z)].
\label{eq:error-list-decoding-eve}
\end{IEEEeqnarray}

\begin{thm}
\label{thm:converse-partition-list-decoding}
Consider an $(M,\error,\secrecy)$ partition code $(\setC,\pi, P_{X|W})$. 
Let $P_{WZ} $ be the distribution defined by the Markov chain $W \to X\to Z$ where $W$ is equiprobable over~$\setM$. 
Then,  for every $Q_Z$, we have
\begin{IEEEeqnarray}{rCl}
d(P_{WZ}, P_WQ_Z) \geq 1-\error_{\mathrm{ld}} - \frac{L}{M} 
\end{IEEEeqnarray}
where $L$ and $\error_{\mathrm{ld}}$ are defined in~\eqref{eq:list-size-eve} and~\eqref{eq:error-list-decoding-eve}, respectively. 
\end{thm}
\ifthenelse{\boolean{conf}}{}{
\begin{IEEEproof}
See Appendix~\ref{app:proof-pa-list}.
\end{IEEEproof}}
By Theorem~\ref{thm:converse-partition-list-decoding}, every achievability bound for the list decoding at the eavesdropper yields a converse bound on the secrecy rate of the wiretap channel.

\subsection{Asymptotic Analysis}

\subsubsection{DM-WTC}

We shall use the following notation
\begin{IEEEeqnarray}{c}
I(P_X,P_{Y|X}) \define I(X;Y)   \\
V(P_X, P_{Y|X}) \define  \! \sum\limits_{x\in\setX}  \! P_X(x)\bigg(\! \sum\limits_{y\in \setY} P_{Y|X}(y|x)   \log^2\! \frac{P_{Y|X}(y|x)}{P_Y (y)}  \notag\\
 \qquad\qquad\qquad\quad - \,D(P_{Y|X=x}\| P_Y)^2 \bigg)\\
\tilde{I}(P_X, P_{YZ|X}) \define  I(X;Y|Z)
\end{IEEEeqnarray}
and
\begin{IEEEeqnarray}{rCl}
\IEEEeqnarraymulticol{3}{l}{
\tilde{V}(P_X,P_{YZ|X})}\notag\\
  &\define&   \sum\limits_{x\in\setX}   P_X(x)\bigg(\! \sum\limits_{y,z }P_{ZY|X}(y,z|x)  \log^2\! \frac{P_{YZ|X}(y,z|x)}{P_{Z|X}(z|x)P_{Y|Z} (y|z)} \notag\\
 &&  \qquad\qquad\quad\quad  -\, D(P_{YZ|X=x}\| P_{Y|Z}P_{Z|X=x})^2\! \bigg).\IEEEeqnarraynumspace
\end{IEEEeqnarray}

The secrecy capacity of a  general DM-WTC is given by~\cite{csiszar1978-05a}
\begin{IEEEeqnarray}{rCl}
\CS = \max_{P_{VX}} \mathopen{}\Big( I(V;Y) - I(V;Z)\Big) 
\label{eq:secrecy-capacity-dmc}
\end{IEEEeqnarray}
where the maximization is over all probability distributions $P_{VX}$  for which $V\to X\to YZ $ form a Markov chain.  For simplicity, we shall assume that there exists a unique probability distribution $P_{VX}^* =P_{V}^* P_{X|V}^*$   that achieves the maximum in~\eqref{eq:secrecy-capacity-dmc}. Note that if the eavesdropper's channel $P_{Z|X}$  is less capable than the legitimate channel $P_{Y|X}$, then the secrecy capacity reduces to~\eqref{eq:secrecy-capacity-intro}~\cite[Sec.~3.5.1]{bloch11-b}.

 The auxiliary random variable $V$ makes the evaluation of~\eqref{eq:secrecy-capacity-dmc} difficult. An upper bound on~\eqref{eq:secrecy-capacity-dmc} is given by~\cite{hayashi2014-09a}
\begin{equation}
\CS \leq \CS^{\mathrm{u}} \define  \max_{P_X} I(X;Y|Z)  .
\label{eq:ub-dmc-secrecy-capacity}
\end{equation}
 For simplicity, we shall also assume that there exists a unique probability distribution  $\tilde{P}_X^*$   that attains the maximum in~\eqref{eq:ub-dmc-secrecy-capacity}, and that $\tilde{V}(\tilde{P}_X^*, P_{YZ|X})>0$. 
Note that, the bound~\eqref{eq:ub-dmc-secrecy-capacity} is tight (i.e., $\CS = \CS^{\mathrm{u}}$) if the wiretap channel  is physically degraded~\cite[Def.~3.8]{bloch11-b}.

\begin{thm}
\label{thm:dm-wtc-asy}
For a DM-WTC $P_{YZ|X}$, we have 
\begin{equation}
R^*(n,\error,\secrecy) \geq \CS - \!\sqrt{\frac{V_1}{n}}Q^{-1}(\error) - \!\sqrt{\frac{V_2}{n}}Q^{-1}(\secrecy) + \bigO\lefto(\frac{\log n}{n}\right)
\label{eq:thm-ach-expansion}
\end{equation}
and 
\begin{equation}
R^*(n,\error,\secrecy) \leq \CS^{\mathrm{u}}- \sqrt{\frac{V_c}{n}}Q^{-1}(\error+\secrecy) +\bigO\lefto(\frac{\log n}{n}\right).
\label{eq:eq:thm-conv-expansion}
\end{equation}
Here, $Q^{-1}(\cdot)$ is the inverse of the Gaussian $Q$-function $Q(x) \define \int\nolimits_{x}^{\infty} \frac{1}{\sqrt{2\pi}} e^{-t^2/2} dt$ and
\begin{IEEEeqnarray}{rCl}
V_1 &\define& V(P_V^*, P_{Y|X}\circ P_{X|V}^*)\\
V_2 &\define& V(P_V^*, P_{Z|X}\circ P_{X|V}^*)\\
V_c &\define& \tilde{V}(\tilde{P}_X^*, P_{YZ|X}).
\end{IEEEeqnarray}

\end{thm}

\ifthenelse{\boolean{conf}}{}{
\begin{IEEEproof}
See Appendix~\ref{app:proof-asy-dmc}.
\end{IEEEproof}}

\begin{rem}
The result~\eqref{eq:thm-ach-expansion} is tighter than the achievable second-order coding rate in~\cite{yassaee2013-07a} obtained by using output statistics of random binning, and tighter than the one in~\cite{tan2012-11a} obtained via channel resolvability. 
The latter two approaches use a random coding argument  and bound the average error probability and average information leakage averaged over all random codebooks separately. They then invoke  Markov's inequality to show the existence of a code that satisfies \emph{simultaneously} the reliability and secrecy constraint. 
The use of Markov's inequality introduces a penalty to the second-order coding rate, which corresponds to the gap between~\eqref{eq:thm-ach-expansion} and~\cite[Eq.~(23)]{yassaee2013-07a}.
In contrast,  our result shows that \emph{every} code that satisfies the reliability constraint can be modified to satisfy the secrecy constraint, thereby avoiding the use of Markov's inequality. This demonstrates the advantage of the privacy amplification technique for wiretap channels in the finite blocklength regime.
\end{rem}

\subsubsection{Gaussian wiretap channel}
Consider the Gaussian wiretap channel   
\begin{IEEEeqnarray}{rCl}
Y_i &=& X_i + U_{i},\quad  Z_i = X_i + \tilde{U}_i,\quad i=1,\ldots,n \label{eq:Gaussian-wiretap}
\end{IEEEeqnarray}
where $\{U_{i}\}$ are independent and identically distributed (i.i.d.)  $\mathcal{N}(0,N_1)$ distributed, and $\{\tilde{U}_i\}$ are i.i.d. $\mathcal{N}(0, N_2)$ distributed. 
Without loss of generality, we assume that $N_2 > N_1$ (otherwise the secrecy capacity is zero).
 Furthermore, we assume that each codeword~$x^n$ satisfies the power constraint
\begin{IEEEeqnarray}{rCl}
 \|x^n\|^2 \leq n P.  
 \end{IEEEeqnarray}
 %

\begin{thm}
\label{thm:gaussian-asy}
For the Gaussian wiretap channel~\eqref{eq:Gaussian-wiretap}, we have 
\begin{equation}
R^*(n,\error,\secrecy) \geq \CS -\sqrt{\frac{V_1}{n}}Q^{-1}(\error) - \sqrt{\frac{V_2}{n}}Q^{-1}(\secrecy) + \bigO\lefto(\frac{\log n}{n}\right)
\label{eq:thm-ach-expansion-awgn}
\end{equation}
and 
\begin{equation}
R^*(n,\error,\secrecy) \leq  \CS -\sqrt{\frac{V_c}{n}}Q^{-1}(\error+\secrecy) +\bigO\lefto(\frac{\log n}{n}\right)
\label{eq:thm-conv-expansion-awgn}
\end{equation}
where 
\begin{IEEEeqnarray}{rCl}
\CS &=&\frac{1}{2}\log\lefto(1+\frac{P}{N_1}\right) - \frac{1}{2}\log\lefto(1+\frac{P}{N_2}\right)\\
V_i &=& \frac{\log^2 e}{2}\frac{P^2+2PN_i}{(P+N_i)^2}, \quad i\in\{1,2\}\\
V_c&=& V_1+V_2 - \frac{PN_1}{P+N_1} \left(\frac{1}{N_2} + \frac{1}{P+N_2}\right)\log^2 e . \label{eq:def-disp-gau-c} \IEEEeqnarraynumspace
\end{IEEEeqnarray}
\end{thm}
\ifthenelse{\boolean{conf}}{}{
\begin{IEEEproof}
See Appendix~\ref{app:proof-asy-gaussian}.
\end{IEEEproof}
}

\subsection{Numerical Results and Discussions}
\ifthenelse{\boolean{conf}}{
\begin{figure}
\begin{center}
\includegraphics[scale=0.65]{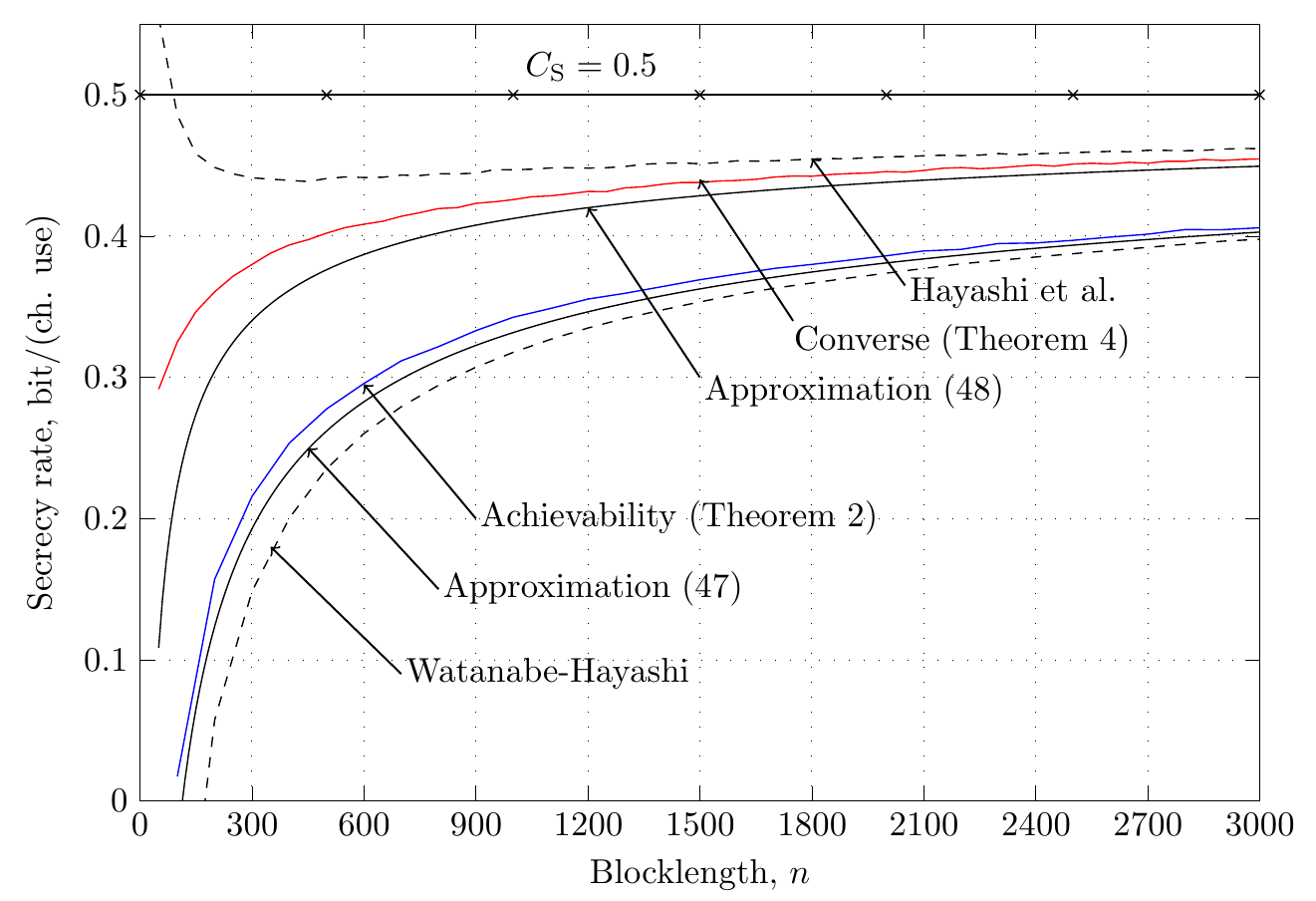}
\caption{Secrecy rate for Gaussian wiretap channel with $P/{N_1} = 3$\, dB, $P/{N_2} = -3$ dB, $\error=\delta=10^{-3}$. \label{fig:gaussian}  }
\end{center}
\end{figure}

}{
\begin{figure}
\begin{center}
\includegraphics[scale=1]{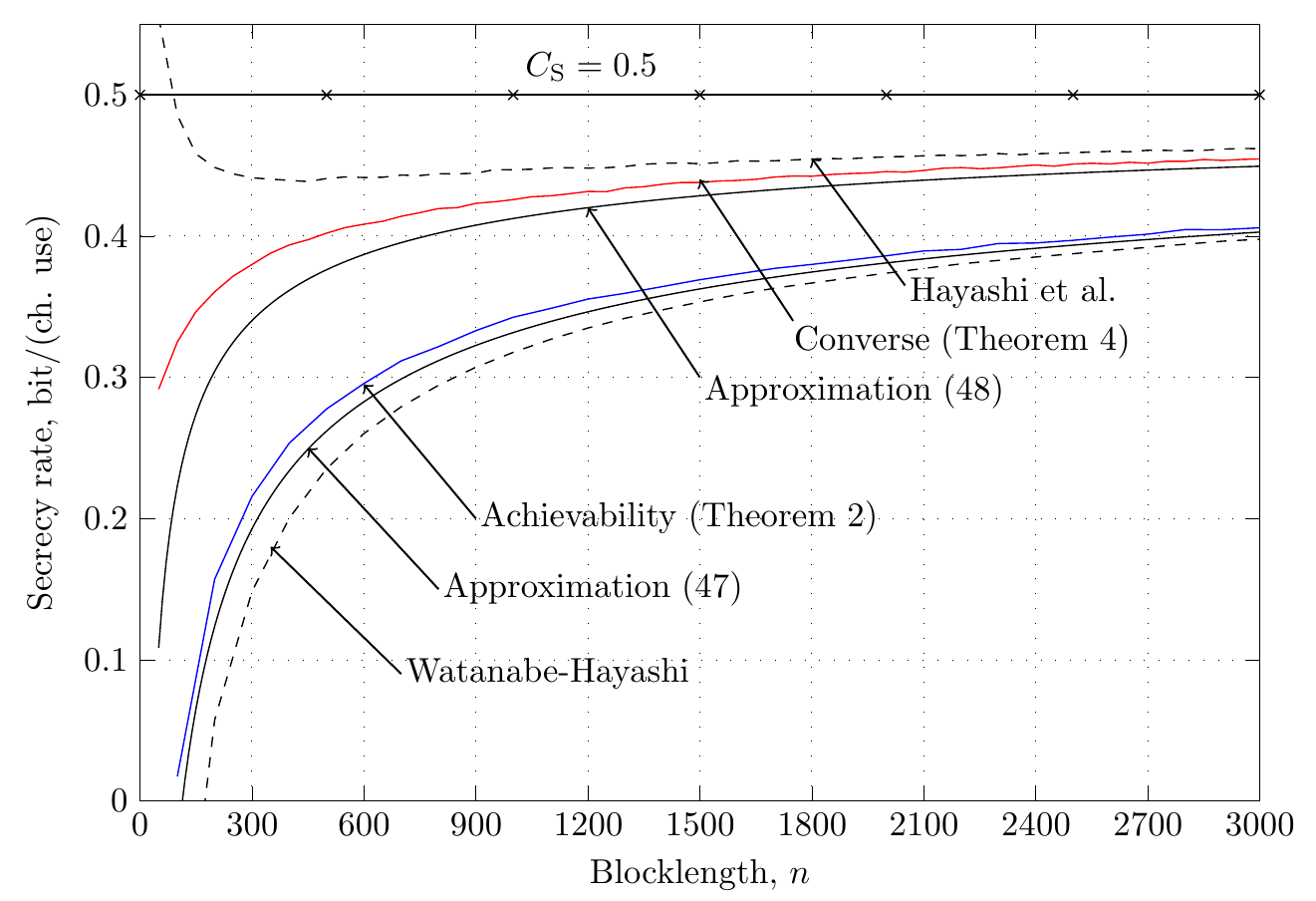}
\caption{Secrecy rate for Gaussian wiretap channel with $P/{N_1} = 3$\, dB, $P/{N_2} = -3$ dB, $\error=\delta=10^{-3}$. \label{fig:gaussian}  }
\end{center}
\end{figure}
}

In this section, we compare the bounds proposed in this paper with existing bounds in~\cite{hayashi2014-09a,watanabe2013-07a} and with the approximations provided in Theorem~\ref{thm:gaussian-asy} for a Gaussian wiretap channel (with the $\bigO(\cdot)$ terms omitted). The results are shown in Fig.~\ref{fig:gaussian}. The bound labeled by ``Hayashi et al.'' is~\cite[Th.~6]{hayashi2014-09a} (see also~\eqref{eq:hayashi-bound-wtc}), and the one labeled by ``Watanabe-Hayashi'' is obtained by combining the privacy amplification bound~\cite[Cor.~2]{watanabe2013-07a} with Shannon's channel coding bound~\cite{shannon59} (which is the tightest channel coding achievability bound for Gaussian channels~\cite[Sec.~III.J-4]{polyanskiy10-05}).\footnote{The numerical routine used to compute Shannon's channel coding bound is available at https://github.com/yp-mit/spectre}

Several observations are in order. First of all, both our achievability and converse bounds are uniformly better than the ones in~\cite{watanabe2013-07a,hayashi2014-09a}. Secondly, the expansions~\eqref{eq:thm-ach-expansion-awgn}  and~\eqref{eq:thm-conv-expansion-awgn} provide reasonable approximations for the  bounds in Theorems~\ref{thm:ach-wiretap} and~\ref{thm:converse-wei}.  
Last but not least, there is a nontrivial gap between our achievability and converse bounds (which can also be inferred from the approximations~\eqref{eq:thm-ach-expansion-awgn}  and~\eqref{eq:thm-conv-expansion-awgn}). 
Narrowing down this  gap seems to require more sophisticated tools than the one used in this paper, and are left for future investigations. On a related note, it is also interesting to study whether the converse bounds in Theorems~\ref{thm:partition-converse-beta-beta} and~\ref{thm:converse-partition-list-decoding} for   partition codes and uniform-partition codes   lead to a tighter second-order coding rate characterization than the ones in~\eqref{eq:eq:thm-conv-expansion} and~\eqref{eq:thm-conv-expansion-awgn}.

\ifthenelse{\boolean{conf}}{}{

\begin{appendix}

\subsection{Proof of Lemma~\ref{lemma:privacy-amplification}}

\label{app:proof-of-lemma-privacy-amp}
Let $\pi_0:  \setC \to \{1,\ldots, M\}$ be an arbitrary function  that satisfies $\big|\pi_0^{-1}(m)\big| = K$ for every $m\in \{1,\ldots,M\}$. Let $\setS$ denote the permutation group of $\setC$. 
Furthermore, let $\setF\define\{\pi_0 \circ \sigma: \,\sigma \in \setS\} $. It is not difficult to check that for every  $ x_j \neq x_k \in \setC $ 
\begin{IEEEeqnarray}{rCl}
\sum\limits_{\pi \in \setF} \indfun{ \pi(x_j) =\pi(x_k)} \leq \frac{|\setF|}{M}.
\label{eq:universal-hash-condition}
\end{IEEEeqnarray}
Furthermore, for every function $\pi\in \setF$ and every $m\in\{1,\ldots, M\}$, $|\pi^{-1}(m)| = K$, which implies that $\pi(X)$ is uniformly distributed over $\{1,\ldots, M\}$.
Let $\Pi$ be uniformly distributed over~$\setF$. Then, $\Pi$ is a \emph{universal}$_2$ function~\cite{Carter79-02a}. 
%


%
%

Using the triangular inequality and the left-over hash lemma~\cite[Lemma~5.4.3]{renner2005-09a} (see also~\cite[Sec. II.B]{watanabe2013-07a}), we obtain that for every  nonnegative measure $R_{XZ}$ (not necessary a probability measure) and every probability distribution $Q_Z$
\begin{IEEEeqnarray}{rCl}
\IEEEeqnarraymulticol{3}{l}{
 \Ex{\Pi}{d(P_{\Pi(X) Z} , P_{\Pi(X)}P_Z) } }\notag\\
  \quad &\leq&   \|P_{XZ} -R_{XZ} \|_1 + \frac{1}{2}\sqrt{M \exp( - H_{2}(R_{XZ}|Q_Z) ) } \label{eq:rewrite-egamma-0}  
\end{IEEEeqnarray}
where $H_2(R_{XZ}|Q_Z) \define -\log \sum_{x,z} R_{XZ}(x,z)^2/Q_Z(z)$ denotes the conditional R\'{e}nyi entropy of order 2 relative to $Q_Z$.

We next minimize the RHS of~\eqref{eq:rewrite-egamma-0}   over all $R_{XZ}$. Consider the following optimization problem 
\begin{IEEEeqnarray}{rCl}
\min_{R_{XZ}: \| P_{XZ} -R_{XZ}  \|_1\leq \tilde{\epsilon} } \sum_{x,z} \frac{R_{XZ}(x,z)^2}{Q_{Z}(z)}
\label{eq:optimizing-h2-min}
\end{IEEEeqnarray}
where $\tilde{\epsilon}\in [0,2]$.
It is not difficult to see that the optimal  $R^*_{XZ}$ must satisfy $R_{XZ}^*(x,z) \leq P_{XZ}(x,z)$ for every pair $(x,z)$, i.e.,~\eqref{eq:optimizing-h2-min} is equivalent to 
\begin{IEEEeqnarray}{rCl}
\min_{R_{XZ}: \| P_{XZ} -R_{XZ}  \|_1\leq \tilde{\epsilon}, \, R_{XZ} \leq P_{XZ} } \sum_{x,z} \frac{R_{XZ}(x,z)^2}{Q_{Z}(z)}.
\label{eq:optimizing-h2-min-2}
\end{IEEEeqnarray}
 Indeed, suppose on the contrary that there exists $(x_0,z_0)$ which satisfy $R^*_{XZ}(x_0,z_0) > P_{XZ}(x_0, z_0)$. Since the objective function in~\eqref{eq:optimizing-h2-min} is monotonically increasing with $R_{XZ}(x_0,z_0)$,  we can further decrease it by setting $R^*_{XZ}(x_0,z_0) = P_{XZ}(x_0,z_0)$ without violating the constraint  in~\eqref{eq:optimizing-h2-min}. But this contradicts the assumption that $R^*_{X,Z}$ is the optimizer of~\eqref{eq:optimizing-h2-min}. Therefore, we must have $R^*_{X,Z} \leq P_{XZ}$. 
 Observe that the problem~\eqref{eq:optimizing-h2-min-2} is a convex optimization problem. Hence, by the Karush-Kuhn-Tucker  (KKT) condition~\cite[Sec.~5.5.3]{boyd04}, the optimizer of~\eqref{eq:optimizing-h2-min-2} must take the form
\begin{IEEEeqnarray}{rCl}
R^*_{XZ}(x,z) \define   \begin{cases}
P_{XZ}(x,z) & \imath(x;z)  \leq \log \gamma \\
\gamma P_{X}(x)Q_Z(z)& \text{otherwise}\\
\end{cases}
\label{eq:def-rxz}
\end{IEEEeqnarray}  
where $\gamma$ is chosen such that $\|P_{XZ} -R_{XZ}^*\|_1 = \tilde{\epsilon}$.

Evaluating the RHS of~\eqref{eq:rewrite-egamma-0} using~\eqref{eq:def-rxz}, we obtain
\begin{IEEEeqnarray}{rCl}
\|P_{XZ} -R^*_{XZ}\|_1 = E_{\gamma}(P_{XZ}, P_XQ_Z).
\label{eq:evaluate-d1-egamma}
\end{IEEEeqnarray}
The second term on the RHS of~\eqref{eq:rewrite-egamma-0} can be evaluated as follows:
\begin{IEEEeqnarray}{rCl}
\IEEEeqnarraymulticol{3}{l}{
\exp(-H_2(R^*_{XZ}|Q_Z)) }\notag\\
 &=& \sum\limits_{x,z} \frac{R^*_{XZ}(x,z)^2}{Q_Z(z)}\\
&=& \frac{1}{KM} \sum\limits_{x,z} P_{XZ}(x,z) \exp(\imath(x;z)) \indfun{\imath (x;z) \leq \log \gamma } \notag\\
&&+\,  \frac{\gamma^2}{KM} P_XQ_Z[\imath(X;Z) \geq \log \gamma]\\
&=& \frac{\gamma}{KM} \Ex{P_{XZ}}{ \exp(-|\imath(X;Z) -\log \gamma|)} \notag\\
&&-\, \frac{\gamma}{KM} \Ex{P_{XZ}}{ \exp(-|\imath(X;Z) -\log\gamma|) \indfun{\imath(X;Z) \geq \log\gamma}}\notag\\
&&+\,  \frac{\gamma^2}{KM} P_XQ_Z[\imath(X;Z) \geq \log \gamma]\\
&=&   \frac{\gamma}{KM} \Ex{}{\exp\mathopen{}\big(\! - \! |\imath(X;Z)-\log \gamma |\big)}.\label{eq:evaluate-expe-term-pa}
\end{IEEEeqnarray}
Substituting~\eqref{eq:evaluate-d1-egamma} and~\eqref{eq:evaluate-expe-term-pa} into~\eqref{eq:rewrite-egamma-0}, we conclude that for every $\gamma>0$
\begin{IEEEeqnarray}{rCl}
&& \Ex{\Pi}{d(P_{\Pi(X) Z} , P_{\Pi(X)}P_Z) } 
  \leq    E_{\gamma}(P_{XZ} , P_XQ_Z) \notag\\
  &&\qquad\qquad\qquad\quad+ \frac{1}{2}\sqrt{ \frac{\gamma}{K}  \Ex{}{\exp\mathopen{}\big(\! - \! |\imath(X;Z)-\log \gamma |\big)} } \label{eq:rewrite-egamma-3} .\IEEEeqnarraynumspace
\end{IEEEeqnarray}
The inequality~\eqref{eq:rewrite-egamma-3}  implies that there exists a $\pi\in\setF$ for which~\eqref{eq:privacy-amplification-lemma} holds.

\subsection{Proof of Theorem~\ref{thm:ach-wiretap}}

\label{app:proof-ach-thm}

We first generate a random codebook $\{X_1,\ldots,X_{KM}\}$ of size $KM$, where the codewords $X_i$ are independent and identically distributed (i.i.d.) according to $P_X$. 
By the random coding union (RCU) bound~\cite[Th.~16]{polyanskiy10-05} and the dependence testing (DT) bound~\cite[Th.~17, Eq.~(79)]{polyanskiy10-05}  the average error probability averaged over all codebooks are upper-bounded by the RHS of~\eqref{eq:min-rcu-beta-bound}. Hence, there exists at least one code $\setC=\{x_1,\ldots,x_{KM}\}$ whose error probability on the channel $P_{Y|X}$ satisfies~\eqref{eq:min-rcu-beta-bound}. 

Next, we construct a secrecy code for the wiretap channel $P_{YZ|X}$ from this code $\setC$ using Lemma~\ref{lemma:privacy-amplification}. Let $P_{X}^{\setC}$ denote the uniform distribution over $\setC$, and let $X\sim P_{X}^{\setC}$. 
Then,  by Lemma~\ref{lemma:privacy-amplification}, there exists a function $\pi:\setC \to \setM$ such that $\pi(X)$
is uniformly distributed over $\setM$, and that for every $\gamma>0$ and every $Q_Z$
\begin{IEEEeqnarray}{rCl}
\IEEEeqnarraymulticol{3}{l}{
d(P_{\pi(X)Z} , P_{\pi(X)}P_Z )
 }\notag\\  
\quad  &\leq&  E_{\gamma}(P_{XZ}^{\setC}, P_X^{\setC}Q_Z) \notag\\
 && +\, \frac{1}{2}\sqrt{\frac{\gamma}{K} \Ex{P_{XZ}^{\setC}}{\exp\mathopen{}\big(\!-\!|\imath(X;Z) -\log \gamma |\big)}}  \\
&\leq&  \sup_{x\in\setA}  E_{\gamma}(P_{Z|X=x} ,Q_Z) \notag\\
&&+\, \sup_{x\in \setA} \frac{1}{2}\sqrt{\frac{\gamma}{K} \Ex{}{ \exp\lefto( - \big|\imath(x;Z)  - \log\gamma \big| \right)}  } . \IEEEeqnarraynumspace
\label{eq:privacy-amplification-lemma-application}
\end{IEEEeqnarray}
Here, the last step follows because the map $P_X\mapsto E_{\gamma}(P_{XZ},P_XQ_Z)$ is linear.

Consider now a random encoder $P_{X|W}$, where for each message $m\in\setM$ it picks a codeword~$X$ from the set $\pi^{-1}(m)$ uniformly at random. The decoder first decodes the codeword~$\hat{X}$   and then set $\hat{W} = \pi(\hat{X})$. 
By construction, the resulting is a uniform-partition code whose error probability is upper-bounded by~\eqref{eq:min-rcu-beta-bound}. Furthermore, we have 
\begin{IEEEeqnarray}{rCl}
P_{W,Z} (m,z)&=& \frac{1}{M}\cdot \frac{1}{|\pi^{-1}(m)|}\sum\limits_{x\in\pi^{-1}(m)} P_{Z|X}(z|x)\\
&=& \frac{1}{MK} \sum\limits_{x\in\pi^{-1}(m)} P_{Z|X}(z|x)\\
&=& P_{\pi(X)Z}(m,z) . 
\label{eq:coincide-distribution}
\end{IEEEeqnarray}
From~\eqref{eq:privacy-amplification-lemma-application} and~\eqref{eq:coincide-distribution}, we conclude that the resulting code also satisfies the secrecy condition~\eqref{eq:thm-secrecy-bound}.

\subsection{Proof of Theorem~\ref{thm:converse-wei}}
\label{app:proof-thm-general-converse}
\begin{figure}
\begin{center}
\includegraphics[scale=0.85]{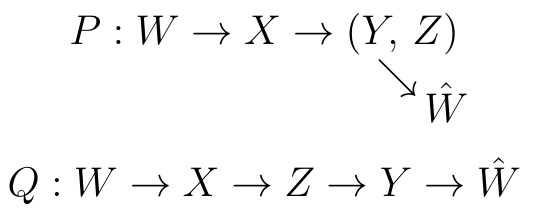}
\end{center}
\caption{A graphical illustration of the probability distributions $P_{WXYZ\hat{W}}$ and $Q_{WXYZ\hat{W}} $. \label{fig:markov}}
\end{figure}
Fix an arbitrary $(M,\error,\secrecy)$ secrecy code  and let $P_{WXYZ\hat{W}} \define P_W P_{X|W} P_{YZ|X} P_{\hat{W} |Y}$ be the joint distribution induced by the code. And let $Q_{WXYZ\hat{W}} \define P_W P_{X|W}P_{Z|X} Q_{Y|Z} P_{\hat{W}|Y } $ (see Fig.~\ref{fig:markov} for a graphical illustration). 
Furthermore, let $P_{T|W  \hat{W}Z}  : \setM^2 \times \setZ \mapsto\{0,1\}$ be defined as
\begin{IEEEeqnarray}{rCl}
T(m,\hat{m},z)=\indfun{ m = \hat{m} , P_{W|Z}(m|z) \leq 1/(\eta M) } \IEEEeqnarraynumspace
\end{IEEEeqnarray}
where $\eta\in(0,1)$.

As in the  meta-converse bound~\cite[Th.~26]{polyanskiy10-05} and in~\cite{tyagi2015-05a}, the idea is to   use $T$ as a suboptimal test between $P_{W\hat{W}Z}$ and $Q_{W\hat{W}Z}$. 
We have
\begin{IEEEeqnarray}{rCl}
\IEEEeqnarraymulticol{3}{l}{
Q_{W\hat{W} Z} [T=1] } \notag\\
\quad &=& \sum_{m,\hat{m},z} Q_{Z\hat{W}}(z,\hat{m})P_{W|Z}(m|z) \notag\\
&& \quad \quad \cdot\, \indfun{m = \hat{m}, P_{W|Z}(m|z)\leq 1/(M\eta) } \\
&\leq & \frac{1}{ M \eta}\sum\limits_{m,z} Q_{Z\hat{W}}(z,m)  \indfun{ P_{W|Z}(m|z)\leq 1/(M\eta) } \IEEEeqnarraynumspace\\
&\leq & \frac{1}{M\eta}. \label{eq:ub-q-t-1}
\end{IEEEeqnarray}
The probability $P_{W\hat{W}Z}[T=1]$ can be lower-bounded as
\begin{IEEEeqnarray}{rCl}
P_{W\hat{W}Z} [T=1] &\geq& 1-\error - P_{WZ}[P_{W|Z}(W|Z)\geq 1/(M\eta)] \IEEEeqnarraynumspace
\label{eq:upper-bound-P-ac-1}
\end{IEEEeqnarray}
which follows from~\eqref{eq:error-constraint-wtc}.
To further lower-bound the RHS of~\eqref{eq:upper-bound-P-ac-1}, we observe that
\begin{IEEEeqnarray}{rCl}
\delta &\geq& d(P_{WZ},P_WP_Z)\\
 &\geq&  P_{WZ}[P_{W|Z}(W|Z)\geq 1/(M\eta)] \notag\\
&&- \, P_{W}P_{Z}[P_{W|Z}(W|Z)\geq 1/(M\eta)] \IEEEeqnarraynumspace
\label{eq:bound-pwz-id-1}
\end{IEEEeqnarray}
and that
\begin{IEEEeqnarray}{rCl}
\IEEEeqnarraymulticol{3}{l}{
P_{WZ}[P_{W|Z}(W|Z)\geq 1/(M\eta)] }\notag\\
\quad & \geq & \frac{1}{\eta} P_{W}P_{Z}[P_{W|Z}(W|Z)\geq 1/(M\eta)].
\label{eq:bound-pwz-id-2}
\end{IEEEeqnarray}
Combining~\eqref{eq:bound-pwz-id-1} and~\eqref{eq:bound-pwz-id-2}, we obtain
\begin{IEEEeqnarray}{rCl}
P_{WZ}[P_{W|Z}(W|Z)\geq 1/(M\eta)] \leq \frac{\delta}{1-\eta}.
\label{eq:bound-pwz-id-3}
\end{IEEEeqnarray}
Substituting~\eqref{eq:bound-pwz-id-3} into~\eqref{eq:upper-bound-P-ac-1}, and using~\eqref{eq:ub-q-t-1}, we conclude that 
\begin{IEEEeqnarray}{rCl}
\beta_{1-\error-\secrecy/(1-\eta)} ( P_{W\hat{W}Z} , Q_{W\hat{W}Z}) \leq \frac{1}{M\eta}.
\label{eq:ub-beta-mmz}
\end{IEEEeqnarray}
The final bound~\eqref{eq:converse-bound-general} follows by rearranging the terms in~\eqref{eq:ub-beta-mmz}, by the change of variable $\tau =\delta/(1-\eta) -\delta$,  and by    
\begin{equation}
\beta_{\alpha} ( P_{W\hat{W}Z} , Q_{W\hat{W}Z}) \geq \beta_{\alpha} ( P_{XYZ} , P_{XZ}Q_{Y|Z})
\end{equation}
which follows from the data-processing inequality for $\beta_\alpha$.

\subsection{Proof of Lemma~\ref{thm:converse-partition-egamma}}
\label{app:proof-pa-converse-egamma}

Let 
\begin{equation}
\tilde{\imath} (x;z) \define \log \frac{P_{XZ}}{Q_XQ_Z}(x,z).
\end{equation}
By definition, for every $m\in\setM$ and every $\gamma>0$, we have
\begin{IEEEeqnarray}{rCl} 
 d(P_{Z|W=m} , Q_Z) &\geq& P_{Z\given W=m} \lefto[\max_{\bar{x}\in\pi^{-1}(m)} \tilde{\imath} (\bar{x};Z) \geq \gamma \right] \notag \\
 && -\, Q_{Z}\lefto[\max_{\bar{x}\in \pi^{-1} (m)} \tilde{\imath} (\bar{x};Z)  \geq \gamma\right].\IEEEeqnarraynumspace
\label{eq:lower-bound-vd-def}
\end{IEEEeqnarray}
The first term on the RHS of~\eqref{eq:lower-bound-vd-def} can be bounded as 
\begin{IEEEeqnarray}{rCl} 
\IEEEeqnarraymulticol{3}{l}{
P_{Z\given W=m}\lefto[\max_{\bar{x}\in\pi^{-1}(m)} \tilde{\imath} (\bar{x};Z) \geq \gamma \right] }\notag\\
&=&   \sum\limits_{x \in \pi^{-1}(m) } \! P_{X|W=m}(x) P_{Z|X=x} \lefto[ \max_{ \bar{x}\in \pi^{-1}(m)} \tilde{\imath}(\bar{x};Z)\geq \gamma \right] \IEEEeqnarraynumspace\\
&\geq&   \sum\limits_{x \in \pi^{-1}(m) } \!P_{X|W=m}(x) P_{Z|X=x}  \lefto[\tilde{\imath} (x;Z) \geq \gamma \right] \\
&=& P_{XZ|W=m}\lefto[\tilde{\imath}(X;Z) \geq \gamma \right].
\label{eq:lower-bound-vd-def-term1}
\end{IEEEeqnarray}
The second term on the RHS of~\eqref{eq:lower-bound-vd-def} can be bounded using the union bound as follows:
 \begin{IEEEeqnarray}{rCl} 
Q_{Z}\lefto[\max_{x\in \pi^{-1} (m)} \tilde{\imath}(x;Z) \geq \gamma\right] &\leq& \sum\limits_{x\in \pi^{-1}(m)}\!\! Q_{Z}[\tilde{\imath}(x;Z)\geq \gamma] .\IEEEeqnarraynumspace
\label{eq:lower-bound-vd-def-term2}
\end{IEEEeqnarray}
Substituting~\eqref{eq:lower-bound-vd-def-term1} and~\eqref{eq:lower-bound-vd-def-term2} into~\eqref{eq:lower-bound-vd-def},  and averaging both sides of~\eqref{eq:lower-bound-vd-def} over $W\sim P_W$, we obtain 
\begin{IEEEeqnarray}{rCl}
\IEEEeqnarraymulticol{3}{l}{
 d(P_{WZ} , P_WQ_Z) }\notag\\
 \quad &=& \sum\limits_{m\in\{1,\ldots,M\}} \frac{1}{M} d(P_{Z|W=m} , Q_Z)  \\
 &\geq&  \sum\limits_{m\in\{1,\ldots,M\}}  \frac{1}{M}\bigg(P_{XZ|W=m}\lefto[\tilde{\imath}(X;Z) \geq \gamma \right] \notag\\
 && \qquad\qquad\quad - \sum\limits_{x\in \pi^{-1}(m)}Q_{Z}[\tilde{\imath}(x;Z)\geq \gamma]\bigg)\\
 &=& P_{XZ}\lefto[\tilde{\imath}(X;Z) \geq \gamma \right] -  \frac{N}{M} Q_XQ_Z[\tilde{\imath}(X;Z)\geq \gamma]. \IEEEeqnarraynumspace
 \label{eq:final-bound-dwz}
\end{IEEEeqnarray}
The proof is concluded by maximizing the RHS of~\eqref{eq:final-bound-dwz} over~$\gamma$, and by the following identity (see, e.g.,~\cite[Eqs. (16) and (18)]{liu2015-11a})
\begin{IEEEeqnarray}{rCl}
\IEEEeqnarraymulticol{3}{l}{
\sup_{\gamma}\Big\{ P_{XZ}\lefto[\tilde{\imath}(X;Z) \geq \gamma \right] -  \frac{N}{M} Q_{X}Q_{Z}[\tilde{\imath}(X;Z)\geq \gamma]\Big\} } \notag\\
\qquad = E_{N/M}(P_{XZ}, Q_{X} Q_{Z} ).
\end{IEEEeqnarray}

\subsection{Proof of Theorem~\ref{thm:converse-partition-list-decoding}}
\label{app:proof-pa-list}
For every $m\in\setM$, we have
\begin{IEEEeqnarray}{rCl} 
 d(P_{Z|W=m} , Q_Z) &\geq& P_{Z\given W=m} \lefto[\pi^{-1}(m) \cap \setL(Z) \neq \emptyset \right] \notag \\
 && -\, Q_{Z}\lefto[\pi^{-1}(m) \cap \setL(Z) \neq \emptyset\right].\IEEEeqnarraynumspace
\label{eq:lower-bound-vd-def-ld}
\end{IEEEeqnarray}
The first term on the RHS of~\eqref{eq:lower-bound-vd-def-ld} can be bounded as 
\begin{IEEEeqnarray}{rCl} 
\IEEEeqnarraymulticol{3}{l}{
P_{Z\given W=m}\lefto[\pi^{-1}(m) \cap \setL(Z) \neq \emptyset  \right] }\notag\\
&=&   \sum\limits_{x \in \pi^{-1}(m) } \! \!P_{X|W=m}(x) P_{Z|X=x} \lefto[ \pi^{-1}(m) \cap \setL(Z) \neq \emptyset   \right] \IEEEeqnarraynumspace\\
&\geq&   \sum\limits_{x \in \pi^{-1}(m) } \!P_{X|W=m}(x) P_{Z|X=x}  \lefto[ x \in \setL(Z)\right] \\
&=& P_{XZ|W=m}\lefto[X \in \setL_Z\right].
\label{eq:lower-bound-vd-def-term1-ld}
\end{IEEEeqnarray}
For the second term on the RHS of~\eqref{eq:lower-bound-vd-def-ld}, we have 
 \begin{IEEEeqnarray}{rCl} 
 \IEEEeqnarraymulticol{3}{l}{
P_WQ_{Z}\lefto[  \pi^{-1}(W) \cap \setL(Z) \neq \emptyset  \right] }\notag\\
&\leq& \frac{1}{M}   \sum_{m\in\setM}  \sum\limits_{z\in\setZ} Q_Z(z) \sum\limits_{x\in \setL(z)}    \indfun{ x \in \pi^{-1}(m)} \\
&=&  \frac{1}{M} \sum\limits_{z\in\setZ} Q_Z(z) \sum\limits_{x\in \setL(z)}   \sum_{m\in\setM}  \indfun{ x \in \pi^{-1}(m)} \\
& =  & \frac{1}{M} \Ex{Q_Z}{|\setL(Z)|} \\
&\leq & \frac{L}{M}. \IEEEeqnarraynumspace
\label{eq:lower-bound-vd-def-term2-ld}
\end{IEEEeqnarray}
Here, the last step follows from~\eqref{eq:list-size-eve}.
Combining~\eqref{eq:lower-bound-vd-def-term1-ld} and~\eqref{eq:lower-bound-vd-def-term2-ld} with~\eqref{eq:lower-bound-vd-def-ld},  we obtain 
\begin{IEEEeqnarray}{rCl}
\IEEEeqnarraymulticol{3}{l}{
 d(P_{WZ} , P_WQ_Z) }\notag\\
 \quad  &\geq&  \sum\limits_{m\in\setM}  \frac{1}{M}P_{XZ|W=m}\lefto[X\in\setL(Z)\right] -\frac{L}{M}\\
 &=& 1-\error_{\mathrm{ld}} - \frac{L}{M} \IEEEeqnarraynumspace
\end{IEEEeqnarray}
where the last step follows from~\eqref{eq:error-list-decoding-eve}.

\subsection{Proof of Theorem~\ref{thm:dm-wtc-asy}}
\label{app:proof-asy-dmc}

To prove~\eqref{eq:thm-ach-expansion},  we shall use Theorem~\ref{thm:ach-wiretap} with a constant-composition code. 
The reason for using constant-composition codes instead of i.i.d. codes is two-fold. First, for a fixed composition $P_X$ and a properly chosen $Q_{Z^n}$, all codewords $x^n$ of a constant-composition code have the same $E_{\gamma}(P_{Z^n|X^n=x^n }, Q_{Z^n})$. Secondly, constant-composition codes achieve the conditional variances $V_1$ and $V_2$, whereas i.i.d. codes achieve (the slightly bigger) unconditional variances.   

Let $\setP_n$ be the set of types of length-$n$ vectors in $\setX^n$, and let $P_n \in \setP_n$ be the  type that is closest in total variation distance to $P_X^*$. 
Furthermore, let $P_{X^n}$ denote the uniform distribution over the set of codewords of type $P_n$, and let $P_{Y^n} \define P_{Y^n|X^n} \circ P_{X^n}$. 
We evaluate $E_{a}(P_{X^n}P_{Y^n|X^n} , P_{X^n} P_{Y^n})$ for a given $a > 0$ as follows:
\begin{IEEEeqnarray}{rCl}
\IEEEeqnarraymulticol{3}{l}{1-  E_{a}(P_{X^n}P_{Y^n|X^n} , P_{X^n}P_{Y^n}) }\notag\\
&=& \inf_{\gamma_1 >0} \bigg\{ P_{X^n}P_{Y^n|X^n}\lefto[  \frac{P_{Y^n | X^n} }{P_{Y^n}}(X^n,Y^n) \leq \gamma_1 \right] + a P_{X^n}P_{Y^n}\lefto[ \frac{P_{Y^n | X^n} }{P_{Y^n}}(X^n,Y^n) >  \gamma_1 \right]\bigg\}  \label{eq:bound-E-gamma-xy-1}\\
&\leq & P_{X^n}P_{Y^n|X^n}\lefto[ \frac{P_{Y^n | X^n} }{P_{Y^n}}(X^n,Y^n) \leq \sqrt{n} a \right]  + \frac{1}{\sqrt{n}}. \label{eq:bound-E-gamma-xy-2}
\end{IEEEeqnarray}
Here,~\eqref{eq:bound-E-gamma-xy-1} follows from~\eqref{eq:def-e-gamma-2}, and~\eqref{eq:bound-E-gamma-xy-2} follows by relaxing the infimum on the RHS of~\eqref{eq:bound-E-gamma-xy-1} with $\gamma_1 = \sqrt{n} a$ and by applying  the standard change of measure technique to the second term on the RHS of~\eqref{eq:bound-E-gamma-xy-1}. 
Suppose that $V_1 = V(P_X^* ,P_{Y|X})>0$.
Proceeding as in the proof of~\cite[Th.~4.2]{tan14}, we can further upper-bound the RHS of~\eqref{eq:bound-E-gamma-xy-2} by
\begin{IEEEeqnarray}{rCl}
Q\lefto( \frac{n I(P_X^*, P_{Y|X}) - \log a  + \bigO(\log n)}{\sqrt{nV_1}} \right) + \bigO\lefto(\frac{1}{\sqrt{n}}\right).
\label{eq: bd-feinstein-tan}
\end{IEEEeqnarray}
This implies that there exists an
\begin{IEEEeqnarray}{rCl}
 a_n = \exp\mathopen{}\Big( nI(X;Y) -\sqrt{nV_1}Q^{-1}\mathopen{}\big(\error  - \bigO(1/\sqrt{ n})\big) - \bigO(\log n)\Big)
 \end{IEEEeqnarray}
 that satisfies 
\begin{equation}
\error \geq 1- E_{a_n}(P_{X^n} P_{Y^n|X^n} , P_{X^n}P_{Y^n}).
\label{eq: asy-error-eva}
\end{equation}
For the case  $V_1=0$, we have that 
\begin{equation}
V(P_n, P_{Y|X}) = \bigO(1/n)
\label{eq:bound-on-vp}
\end{equation}
which follows because $|P_n -P_X^*| =\bigO(1/n)$ and because  $P \mapsto V(P,P_{Y|X})$ are smooth maps on the interior of the probability simplex on $\setX$.  Proceeding step by step as in the proof of~\cite[Th.~4.2]{tan14},  and using Chebyshev's inequality and~\eqref{eq:bound-on-vp} in place of~\cite[Eq.~(4.53)]{tan14}, we conclude that~\eqref{eq: bd-feinstein-tan} and~\eqref{eq: asy-error-eva} remain to hold if $V_1=0$.

We next evaluate $E_{\gamma}(P_{Z^n|X^n =x^n} , Q_{Z^n})$  for an arbitrary $x^n$ of type $P_{n}$ and for $Q_{Z^n} =(P_{Z|X} \circ P_n)^n$.
In the analysis below, we shall assume that $V_2 = V(P^*_X, P_{Y|X})>0$. The case $V_2=0$ can be handled similarly as in~\eqref{eq:bound-on-vp}.
 Consider the following chain of (in)equalities: 
\begin{IEEEeqnarray}{rCl}
E_{\gamma}(P_{Z^n|X^n =x^n} , Q_{Z^n}) &\leq& P_{Z^n |X^n =x^n}\lefto[\log \frac{P_{Z^n|X^n=x^n} }{ Q_{Z^n}} (Z^n) \geq \log \gamma \right] \\
&= &P_{Z^n |X^n =x^n}\lefto[ \sum\limits_{i=1}^{n} \log \frac{P_{Z|X=x_i}}{(P_{Z|X} \circ P_n)} (Z_i) \geq \log \gamma \right]\\
&\leq& Q\lefto(\frac{ \log \gamma -n I(P_n, P_{Z|X}) }{\sqrt{nV(P_n,P_{Z|X} )} }\right) + \bigO\lefto(\frac{1}{\sqrt{n}}\right)\label{eq:berry-esseen-pzx}\\
&=&  Q\lefto(\frac{ \log \gamma -n I(P_X^*, P_{Z|X}) }{\sqrt{nV(P_X^*,P_{Z|X} )} }\right) + \bigO\lefto(\frac{1}{\sqrt{n}}\right).\label{eq:berry-esseen-pzx-convergence}
\end{IEEEeqnarray}
Here,~\eqref{eq:berry-esseen-pzx} follows from the Berry-Esseen theorem~\cite[Sec. XVI.5]{feller70a} and~\cite[Lemma~46]{polyanskiy10-05}, and~\eqref{eq:berry-esseen-pzx-convergence} follows from because  $|P_n - P_X^*| = \bigO(1/n)$ and because $P\mapsto I(P, P_{Z|X})$ and $P \mapsto V(P,P_{Z|X})$ are smooth maps on the interior of the probability simplex on $\setX$. Setting $\log \gamma_n \define nI(P_X^*, P_{Z|X}) + \sqrt{nV(P_{X}^*,P_{Z|X})}Q^{-1}\mathopen{}\big(\delta - \bigO(1/\sqrt{n}) \big)$ and $K = n \gamma_n$,  and using that $\Ex{}{\exp(-|i(x^n;Z^n) -\log \gamma|)} \leq 1$, we obtain
\begin{IEEEeqnarray}{rCl}
E_{\gamma_n}(P_{Z^n|X^n =x^n} ,Q_{Z^n} )  + \frac{1}{2}\sqrt{\frac{\gamma_n}{K} \Ex{}{\exp(-|i(x^n;Z^n) -\log \gamma|)}  }  \leq \delta. \label{asy-secrecy-eva}
\end{IEEEeqnarray}
Setting $M_n \define  (2a_n+1)/K $, and combining~\eqref{eq: asy-error-eva},~\eqref{asy-secrecy-eva}, and Theorem~\ref{thm:ach-wiretap}, we conclude that
\begin{IEEEeqnarray}{rCl}
R^*(n,\error,\secrecy) &\geq &\frac{1}{n} \log M_n \\
& \geq&  \CS -\sqrt{\frac{V_1}{n}}Q^{-1}(\error) - \sqrt{\frac{V_2}{n}}Q^{-1}(\secrecy) + \bigO\lefto(\frac{\log n}{n}\right).\label{eq:proof-sec-asy-ach-dmc}
\end{IEEEeqnarray}

We next prove the converse bound~\eqref{eq:eq:thm-conv-expansion} using~\eqref{eq:converse-bound-general}. 
%
In order to apply~\eqref{eq:converse-bound-general}, we need to select a $Q_{Y^n \given Z^n}$. 
Before doing so, we remark that in the point-to-point channel coding setting (i.e., without the secrecy constraint), a converse bound is usually proved by reducing a code to a constant-composition subcode (see, e.g.,~\cite{shannon67,polyanskiy10-05}). 
The rationale behind this reduction is that removing all codewords except those of a dominant type  reduces the coding rate by at most $\bigO((\log n)/n)$, but at the same time it  decreases the error probability. 
The $Q_{Y^n}$ can be then chosen as the output distribution induced by the type. 
This reduction argument, however, does not work for the wiretap channel, because it is not clear how removing codewords will affect the secrecy level  $d(P_{WZ},P_WP_Z)$. Instead, we shall choose $Q_{Y^n|Z^n}$ to be a mixture of conditional distributions $P_{Y^n\given Z^n}$ induced by all types in $\setP_n$. We now proceed with the proof.  

For each type $P_X^{(t)}\in \setP_n$, $t=1,\ldots, |\setP_n|$, let $P_{YZ}^{(t)} \define P_{YZ|X} \circ P_X^{(t)}$ and let $P_{Y|Z}^{(t)}$ be the induced conditional distribution.  Furthermore, let 
\begin{IEEEeqnarray}{rCl}
Q_{Y^n|Z^n} (y^n \given z^n)= \frac{1}{|\setP_n|}\sum\limits_{t=1}^{|\setP_n|} \prod\limits_{i=1}^{n} P_{Y|Z}^{(t)}(y_i | z_i).
\label{eq:dmc-q-channel}
 \end{IEEEeqnarray}
Using this conditional distribution in the bound~\eqref{eq:converse-bound-general}, we obtain that, for every $\tau \in(0,1-\error-\delta)$ and every $\gamma>0$,
\begin{IEEEeqnarray}{rCl}
R^*(n,\error,\delta) &\leq&   - \inf_{P_{X^n}}\log \beta_{1-\error -\delta-\tau}  (P_{X^nY^nZ^n},  P_{X^nZ^n}Q_{Y^n\given Z^n}) + \log \frac{\tau+\delta}{\tau} \\
&\leq& \gamma -  \log\lefto( 1-\error-\secrecy-\tau - \sup_{P_{X^n}}
P_{X^nY^nZ^n}\lefto[ \log \frac{P_{X^nY^nZ^n}}{P_{X^nZ^n} Q_{Y^n\given Z^n}} \geq \gamma \right] \right) + \log \frac{\tau+\delta}{\tau} \IEEEeqnarraynumspace \label{eq:bound-max-rate-conv-1}
\end{IEEEeqnarray}
where the second step follows from~\cite[Eq.~(102)]{polyanskiy10-05}.
The probability term on the RHS of~\eqref{eq:bound-max-rate-conv-1} can be evaluated as follows:
\begin{IEEEeqnarray}{rCl}
\IEEEeqnarraymulticol{3}{l}{
P_{X^nY^nZ^n}\lefto[ \log \frac{P_{X^nY^nZ^n}}{P_{X^nZ^n} Q_{Y^n\given Z^n}} \geq \gamma \right] }\notag\\
\,\,&=&  P_{X^nY^nZ^n}\lefto[ \log \frac{P_{Y^nZ^n|X^n}}{ P_{Z^n\given X^n} Q_{Y^n\given Z^n}} \geq \gamma \right]   \label{eq:bound-idcdf-conv-1} \\
&\leq& \sup\limits_{x^n \in \setX^n}  P_{Y^nZ^n\given X^n = x^n}\lefto[ \log \frac{P_{Y^nZ^n|X^n=x^n}}{P_{Z^n\given X^n=x^n}Q_{Y^n\given Z^n}}\geq \gamma \right] . \label{eq:bound-idcdf-conv-2}
\end{IEEEeqnarray}
 For an arbitrary $x^n\in \setX^n$,  let $t_0 \in \{1,\ldots, |\setP_n|\}$ denote the index for which $P_{X}^{(t_0)}$ coincides with the type of $x^n$. 
Using~\eqref{eq:dmc-q-channel} and using that $\log |\setP_n| \leq |\setX|\log (n+1)$, we obtain
\begin{IEEEeqnarray}{rCl}
\IEEEeqnarraymulticol{3}{l}{
P_{Y^nZ^n\given X^n = x^n}\lefto[ \log \frac{P_{Y^nZ^n|X^n=x^n}}{P_{Z^n\given X^n =x^n}Q_{Y^n\given Z^n}}\geq \gamma \right]  }\notag\\
&\leq&  P_{Y^nZ^n\given X^n = x^n}\lefto[ \log \frac{P_{Y^nZ^n|X^n=x^n}}{ P_{Z^n\given X^n = x^n}(P_{Z|Y}^{(t_0)})^n }\geq \gamma -|\setX|\log (n+1)  \right] \\
&=& P_{Y^nZ^n\given X^n = x^n}\lefto[ \sum\limits_{i=1}^{n} \log \frac{P_{YZ|X} (Y_i,Z_i |x_i)}{P_{Z|X}(Z_i|x_i) P_{Z|Y}^{(t_0)}(Y_i|Z_i) } \geq  \gamma - |\setX|\log (n+1) \right].
\label{eq:bound-PYZ}
\end{IEEEeqnarray}

Let now
\begin{IEEEeqnarray}{rCl}
\gamma' (\alpha) \define   \inf \define \left\{\xi:   \sup_{P_{X}^{(t_0)}} P_{Y^nZ^n\given X^n = x^n}\lefto[ \sum\limits_{i=1}^{n} \log \frac{P_{ZY|X} (Y_i,Z_i |x_i)}{P_{Z|X}(Z_i|x_i)P_{Z|Y}^{(t_0)}(Y_i|Z_i) } \geq  \xi   \right] \geq \alpha\right\}. 
\end{IEEEeqnarray}
Following similar steps as in the proof of~\cite[Th.~48]{polyanskiy10-05}, and using that $\tilde{P}_X^*$ is the unique maximizer of~\eqref{eq:ub-dmc-secrecy-capacity} and that $\tilde{V}(\tilde{P}_X^*,P_{YZ|X})>0$, we obtain  
\begin{IEEEeqnarray}{rCl}
\gamma' (\alpha) = n \CS^{\mathrm{u}} -\sqrt{nV_c}Q^{-1}(1-\alpha) + \bigO(\log n).  
\label{eq:expand-gamma'-conv-dmc}
\end{IEEEeqnarray}
Finally, setting~$\gamma = \|\setX\|\log(n+1) + \gamma'(1-\error-\secrecy-\tau)$ and $\tau =1/\sqrt{n}$, and using~\eqref{eq:expand-gamma'-conv-dmc},~\eqref{eq:bound-PYZ}, and~\eqref{eq:bound-idcdf-conv-2}, we conclude the proof of~\eqref{eq:eq:thm-conv-expansion}.

\subsection{Proof of Theorem~\ref{thm:gaussian-asy}}
\label{app:proof-asy-gaussian}

To prove~\eqref{eq:thm-ach-expansion-awgn}, we shall choose $P_{X^n}$ to be the uniform distribution over the power sphere $\setS_n \define \{x^n\in \realset^n:  \|x^n\|^2 = nP\}$ and set $Q_{Z^n} \sim \mathcal{N}(\mathbf{0}, (P+N_2)\matI_n) $. For such a $P_{X^n}$ it was shown in~\cite{tan15-05-a} that there exists an
\begin{equation}
a_n = \exp\mathopen{}\Big(\frac{n}{2}\log(1+P/N_1) -\sqrt{n V_1}Q^{-1}(\error) + \bigO(\log n)\Big)
\end{equation}
that satisfies 
\begin{IEEEeqnarray}{rCl}
\error_{RCU} (a_n)\leq \error. 
\end{IEEEeqnarray}

We next evaluate~\eqref{eq:thm-secrecy-bound}. Due to the spherical symmetry of $\setS_n$ and $Q_{Z^n}$, we have that for every $x^n \in \setS_n$
\begin{IEEEeqnarray}{rCl}
E_{\gamma}(P_{Z^n\given X^n =x^n}, Q_{Z^n}) = E_{\gamma}(P_{Z^n\given X^n =\bar{x}^n} , Q_{Z^n})
\label{eq:eval-egamma-gaussian}
\end{IEEEeqnarray}
where
\begin{equation}
\bar{x}^n\define[\sqrt{P},\ldots, \sqrt{P}].
\label{eq:special-cwd-gaussian}
\end{equation}
 The RHS of~\eqref{eq:eval-egamma-gaussian} can be evaluated 
 \begin{IEEEeqnarray}{rCl}
\IEEEeqnarraymulticol{3}{l}{
E_{\gamma}(P_{Z^n\given X^n =\bar{x}^n} , Q_{Z^n})}\notag\\
 \quad &\leq & P_{Z^n\given X^n= \bar{x}^n}\lefto[ \log \frac{dP_{Z^n\given X^n= \bar{x}^n}}{dQ_{Z^n}} (Z^n) \geq \gamma \right] \\
 &= & Q\lefto(\frac{\gamma -n \log(1+P/N_2)}{\sqrt{nV_2}}\right) + \bigO\lefto( \frac{1}{\sqrt{n}}\right) \IEEEeqnarraynumspace
\end{IEEEeqnarray}
where the last step follows from~\cite[p.~2357]{polyanskiy10-05}. The proof of~\eqref{eq:thm-ach-expansion-awgn} follows by repeating the steps~\eqref{asy-secrecy-eva}--\eqref{eq:proof-sec-asy-ach-dmc}.

To prove the converse bound~\eqref{eq:thm-conv-expansion-awgn}, we assume  that the channel $P_{YZ|X}$ is physically degraded. 
This assumption comes without loss of generality since the maximal coding rate  $R^*(n,\error,\delta)$ depends on the channel law $P_{Y^nZ^n\given X^n}$ only through the marginal transition probabilities $P_{Y^n\given X^n}$ and $P_{Z^n\given X^n}$~\cite[Lemma~3.4]{bloch11-b}. 
We shall use Theorem~\ref{thm:converse-wei} with 
\begin{IEEEeqnarray}{rCl}
Q_{Y^n\given Z^n} = \mathcal{N} \lefto(\frac{P+N_1}{P+N_2} z^n , \frac{(P+N_1)(N_2 -N_1)}{P+N_2} \matI_n\right)
\end{IEEEeqnarray}
which coincides with the marginal conditional distribution $P_{Y^n|Z^n}$ of $P_{X^nY^nZ^n}$  for the case $X^n \sim \mathcal{N}(0, P\matI_n)$.  
Observe that, by data-processing inequality for $\beta_{\alpha}(\cdot,\cdot)$, we have 
\begin{IEEEeqnarray}{rCl}
 \beta_{\alpha} (P_{X^nY^nZ^n}, P_{X^nZ^n}Q_{Y^n|Z^n}) \geq \beta_{\alpha} (P_{X^{n+1}Y^{n+1}Z^{n+1}}, P_{X^{n+1}Z^{n+1}}Q_{Y^{n+1}|Z^{n+1}}). 
 \label{eq:data-pro-beta-conv}
\end{IEEEeqnarray}
By~\eqref{eq:data-pro-beta-conv}, it suffices to consider the case where $P_{X^n}$ is supported on the power sphere $\setS_n$. Furthermore, by the spherical symmetry of $P_{Y^nZ^n|X^n}$, $Q_{Y^n|X^n}$, and~$\setS_n$, and by using~\cite[Lemma~29]{polyanskiy10-05}, we obtain that, for every $P_{X^n}$ supported on $\setS_n$,
\begin{IEEEeqnarray}{rCl}
 \beta_{\alpha} (P_{X^nY^nZ^n}, P_{X^nZ^n}Q_{Y^n|Z^n})  =  \beta_{\alpha} (P_{Y^nZ^n|X^n=\bar{x}^n}, P_{Z^n|X^n=\bar{x}^n}Q_{Y^n|Z^n}) 
 \label{eq:beta-n-n1}
\end{IEEEeqnarray} 
where~$\bar{x}^n$ is defined in~\eqref{eq:special-cwd-gaussian}.

To evaluate the asymptotic behavior of the RHS of~\eqref{eq:beta-n-n1} we observe that, under $P_{Y^nZ^n\given X^n=\bar{x}^n }$, the random variable $\log \frac{dP_{Y^nZ^n\given X^n=\bar{x}^n}}{d(P_{Z^n\given X^n = \bar{x}^n} Q_{Y^n\given Z^n})}(Y^n,Z^n)$ has the same distribution as 
\begin{IEEEeqnarray}{rCl}
n\CS + \frac{\log e}{2} \sum\limits_{i=1}^{n}\left( \frac{(U_i + \bar{U}_i)^2}{N_2} -   \frac{U_i^2}{N_1} + \frac{(\sqrt{P} + U_i)^2}{P+N_1} -\frac{(\sqrt{P} + U_i + \bar{U}_i)^2}{P+N_2}\right)
\label{eq:infod-gaussian}
\end{IEEEeqnarray}
where~$\{\bar{U}_i\}$ are i.i.d. $\mathcal{N}(0, N_2-N_1)$-distributed, and are independent of all other random variables in~\eqref{eq:infod-gaussian}. 
 As in~\cite[Sec.~IV.B]{polyanskiy10-05},  a central limit theorem analysis of~\eqref{eq:infod-gaussian} shows that
 \begin{IEEEeqnarray}{rCl}
-\log \beta_{\alpha} (P_{Y^nZ^n|X^n=\bar{x}^n}, P_{Z^n|X^n=\bar{x}^n}Q_{Y^n|Z^n}) =  n\CS  - \sqrt{n V_c}Q^{-1}(\error+\secrecy+\tau) + \bigO(\log n)
\label{eq:beta-gau-c-exp}
\end{IEEEeqnarray}
where $V_c$ is given in~\eqref{eq:def-disp-gau-c}. Setting $\tau =1/\sqrt{n}$, substituting~\eqref{eq:beta-gau-c-exp} and~\eqref{eq:beta-n-n1} into~\eqref{eq:converse-bound-general}, we conclude the proof of~\eqref{eq:thm-conv-expansion-awgn}.

\end{appendix}

}

\ifthenelse{\boolean{conf}}{

\bibliographystyle{IEEEtran}
\bibliography{IEEEabrv,publishers,confs-jrnls,WeiBib}

}{


}

\end{document}